\documentclass[twocolumn]{article} 
\usepackage[letterpaper, margin=0.7in]{geometry}
\usepackage{amsmath, color}
\usepackage{graphicx}  
\usepackage{dcolumn}   
\usepackage{bm}        
\usepackage{amssymb}   
\usepackage{epstopdf}	 
\usepackage{placeins}
\usepackage{soul,color}
\usepackage[linesnumbered,ruled,vlined]{algorithm2e}

\newcommand{\hb}[1] {{\bm {\hat{#1}} }}

\begin{document}

\title{Vertex stability and topological transitions in vertex models of foams and epithelia}
\author{Meryl A. Spencer \and Zahera Jabeen\thanks{Present address: Department of Mechanical Engineering and Applied Mechanics, University of Pennsylvania, Philadelphia, PA 19104}\ \and David K. Lubensky \\ Department of Physics, University of Michigan, Ann Arbor, MI 48103, USA}
\date{}

\twocolumn[ 
\begin{@twocolumnfalse}
    \maketitle
\begin{abstract}
In computer simulations of dry foams and of epithelial tissues, vertex models are often used to describe the shape and motion of individual cells. Although these models have been widely adopted, relatively little is known about their basic theoretical properties.  For example, while fourfold vertices in real foams are always unstable, it remains unclear whether a simplified vertex model description has the same behavior.  Here, we study vertex stability and the dynamics of T1 topological transitions in vertex models.  We show that, when all edges have the same tension, stationary fourfold vertices in these models do indeed always break up.  In contrast, when tensions are allowed to depend on edge orientation, fourfold vertices can become stable, as is observed in some biological systems.  More generally, our formulation of vertex stability leads to an improved treatment of T1 transitions in simulations and paves the way for studies of more biologically realistic models that couple topological transitions to the dynamics of regulatory proteins. 
\end{abstract}
\end{@twocolumnfalse}
]
\section{Introduction}
\begin{figure*}[!ht]
		\centering
    \includegraphics[width=0.9\textwidth]{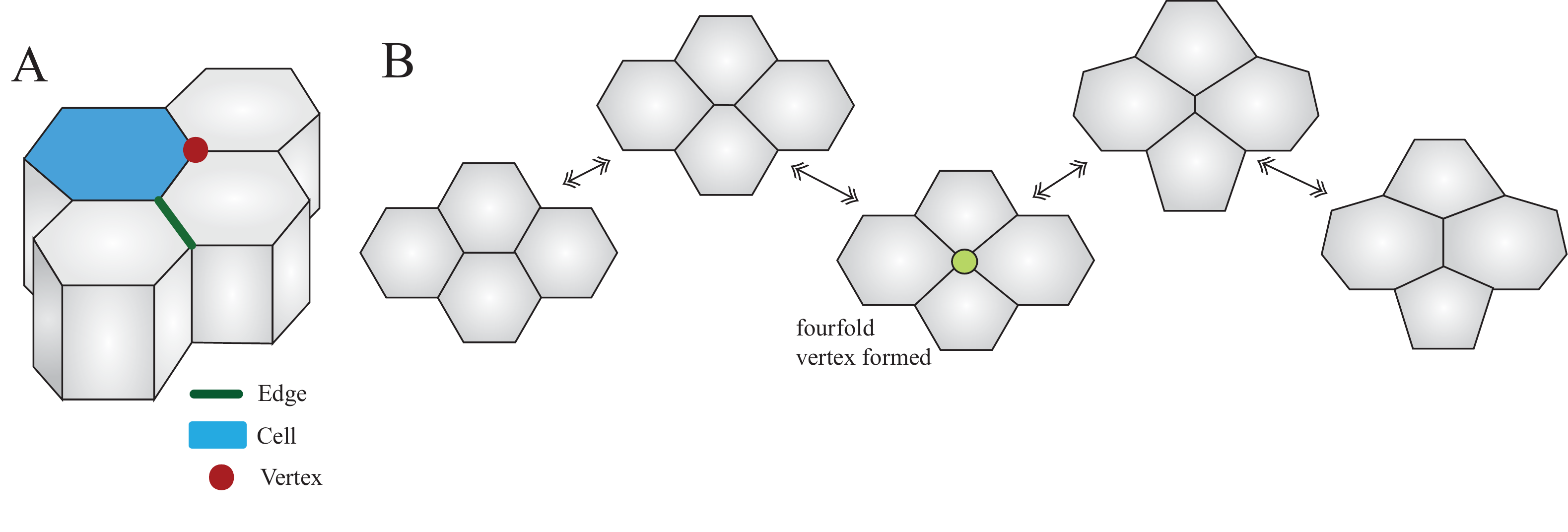}
    \caption{ ${\bf A}$: Cartoon of cells in an epithelial sheet. A single cell is shaded blue. The interface between two cells forms an edge (one edge is highlighted by the bold green line). The red dot indicates a vertex, defined as a point at which three or more cells touch. We treat the epithelium as a two-dimensional sheet, focusing on the level of the adherens junctions near the apical (top) surface. \newline ${\bf B}:$ Cartoon of epithelial cells undergoing a T1 topological transition (viewed from above). An edge shrinks down until a fourfold vertex is formed, then a new edge elongates in a roughly perpendicular direction.  As a result, the cells exchange neighbors, altering the topology of the cell packing. The middle panel shows the moment at which a fourfold vertex (light green dot) appears. The fourfold vertex has four neighboring cells and four neighboring edges and could in principle either be stable or resolve into either of the two different topologies shown to the left and right.}
    \label{fig:intro}
\end{figure*}

From the lining of the gut to the surface of the skin, epithelial tissues are one of the essential building blocks of animal organs.  The motion of epithelial cells over time correspondingly drives many aspects of animal development and morphogenesis, and understanding this movement is thus a central problem in quantitative biology.  Although there has been remarkable progress in identifying and imaging the proteins involved in the development of specific epithelia \cite{Axelrod:PCP,Munro:PCP,Bellaiche:PCP,Eaton:PCP,Mao:PCP,Assémat:PCP},
 it remains a major challenge to translate this molecular knowledge into a higher level picture of how the organization of epithelial cells emerges from local mechanical interactions.  Computational modeling represents an important tool to address this question, and it is hence essential to have well-understood models available to describe epithelia.  Here, we begin to address this need by deriving some general results on the stability of fourfold vertices in a widely-used class of vertex models~\cite{Shvartsman:VM,Maini:VM,Schilling:VM,Juelicher:VM}.

A simple epithelium is a quasi-two-dimensional sheet of cells characterized by strong inter-cellular adhesion \cite{Lecuit:actomyocin,Kemkemer:actomyocin}. This adhesion occurs primarily at a belt of adherens junctions, composed largely of cadherins, which hold the adjacent cell membranes together. Additionally, each cell has a band of contractile cortical acto-myosin running along the inside of the adherens junctions.  The combination of adherens junction and contractile actin ring leads to an effective line tension along cell-cell junctions which the cell can modulate by targeting adhesion molecules or myosin and their regulators.  Thus, for example, the tension can be made to vary as a function of junctional orientation as a result of regulation by the planar cell polarity pathway \cite{Axelrod:PCP,Munro:PCP,Bellaiche:PCP,Eaton:PCP,Mao:PCP,Assémat:PCP} (as seen, e.g., in \textit{Drosophila} germ band extension~\cite{Lecuit:actomyocin}). 

When viewed as a two dimensional sheet of tightly packed cells, the epithelium strongly resembles a dry soap film.  Indeed, interfacial tension plays a central role in the physics of both systems, and foam-inspired models are thus frequently used to describe epithelia.  The standard model for the mechanics of a dry foam, which we here refer to as the \textit{Plateau model}, goes back to the work of Plateau in the 1800s \cite{Hutzler:Proof}. It posits that the final shape of a group of bubbles is determined by minimizing a surface tension energy proportional to the total bubble surface area (in 3 dimensions) or the total length of the interfaces between bubbles (in two dimensions).

Many recent computational descriptions of epithelia have been based on so-called vertex models \cite{Shvartsman:VM,Maini:VM,Schilling:VM,Juelicher:VM}, a class of simplified variants of the Plateau model that have been applied to systems including epithelia, foams, and metal grains \cite{Shvartsman:VM,Honda:Comp,Honda:Checkerb,KAWASAKI:history,Frost:Grain}.  The two models share the basic feature of an energy that grows with the total length of the bubble-bubble or cell-cell interfaces.  They differ in that, whereas the Plateau model allows the interfaces between bubbles or cells to take arbitrary shapes, vertex models impose that (in two dimensions) these interfaces must always be straight lines, which we refer to as \textit{edges}. (Extensions which allow for curved edges \cite{Ishimoto:curved,Lubensky:2012,Sherrard:curved,Chen:curved} and for three-dimensional cells \cite{Hannezo:3D} have been proposed but are beyond the scope of this paper.) The major degrees of freedom in vertex models are then the positions of the \textit{vertices} where three or more cells meet and which are joined by edges to form polygonal cells (fig. \ref{fig:intro}A).  

A fourfold vertex occurs whenever a vertex has four neighboring cells and edges as opposed to the much more common three. Fourfold vertices generally resolve into two threefold vertices by pairing the edges of the fourfold vertex and growing a new edge between them. There are two different ways to pair the edges, resulting in two different final cell arrangements (fig. \ref{fig:intro}B). Cellular rearrangements that switch between these two topologies, through the intermediate of a fourfold vertex, allow the epithelial sheet to change shape and enable cells of specific types to find their correct location and morphology. Indeed, this process, known as a T1 transition, has been shown to play a central role in morphogenetic movements like tissue elongation \cite{Aigouy:T1example,Zallen:T1example,Rauzi:T1}. Though fourfold vertices usually break up, tissues in which fourfold vertices remain stable over a relatively long timescale have also recently been observed \cite{DiNardo:4fold,Bellaïche:4fold,Maini:4fold,Zallen:Square,Harding:review}.

Although vertex models (proposed by Honda in the 1980s \cite{Honda:Comp,Honda:OrigVM,Honda:VMenergy}) clearly ignore many features of real cell shape, they are thought to capture the essential physics when cells are close to polygonal, and they have been applied successfully to study many features of epithelial morphogenesis \cite{Shvartsman:VM,Juelicher:VM,Bellaiche:VM,Reuter:VM}.  Moreover, they have the advantage of being both simple and straightforwardly extensible to include effects ranging from the dynamics of proteins localized at the edges to buckling into the third dimension \cite{Lubensky:2012,Reuter:VM,Du:VM}. Despite their increasing popularity, however, some of these models' fundamental theoretical properties are poorly understood \cite{Staple:VM,Bi:VM}.  Most notably, in the Plateau model of dry foams which inspired vertex models, fourfold vertices (fig. \ref{fig:intro}B) are always unstable, breaking up into two threefold vertices \cite{Hutzler:Proof}.  Because vertex models demand that cell-cell junctions remain straight, cell pressure plays a somewhat different role in them from their role in the Plateau model (where edges can take on any shape), and the standard arguments leading to this instability cannot be taken over directly from the Plateau model.  It is thus unclear whether the instability is likewise always present in vertex models. Here, we show that the vertex model does not allow for stable fourfold vertices at mechanical equilibrium when all edges have the same tension. In contrast, we find that introducing a simple dependence of tensions on edge orientation is sufficient to stabilize fourfold vertices. This result may help to explain the observation of long-lived fourfold vertices in some biological systems 
\cite{Honda:Checkerb,Aigouy:T1example,DiNardo:4fold,Bellaïche:4fold,Maini:4fold,Zallen:Square,Harding:review,Sakurai:leg}. Moreover, our examination of the dynamics of fourfold vertices suggests an improved algorithm for treating T1 transitions in simulations which removes the potential for spurious oscillations and incorrect resolutions present in some prior \textit{ad hoc} approaches. This procedure will be especially useful as we develop more complex models of epithelia that couple cell shape and the dynamics of junctional proteins \cite{Lubensky:2012}.  

In the remainder of this paper, we investigate the stability of fourfold vertices and dynamics near topological transitions in vertex models. We begin with a full description of the model, and we then proceed to develop equations describing the dynamics near fourfold vertices. In section \ref{sec:stab}, we state the conditions under which a fourfold vertex is stable. The subsequent two sections then show that it is impossible to satisfy all of the stability conditions simultaneously for stationary vertices with equal tensions, demonstrating that the model does not admit stable fourfold states in this case. In section \ref{sec:stabelex}, we argue that stable fourfold vertices do become possible when the assumptions of mechanical equilibrium or of equal tensions are relaxed, potentially shedding light on why fourfold vertices are observed in some biological systems. We conclude by touching on the implications of our results for the design of algorithms to simulate vertex models.  

\section{The vertex model}
\label{sec:VModel}
\begin{figure}[]
		\centering
    \includegraphics[width=0.3\textwidth]{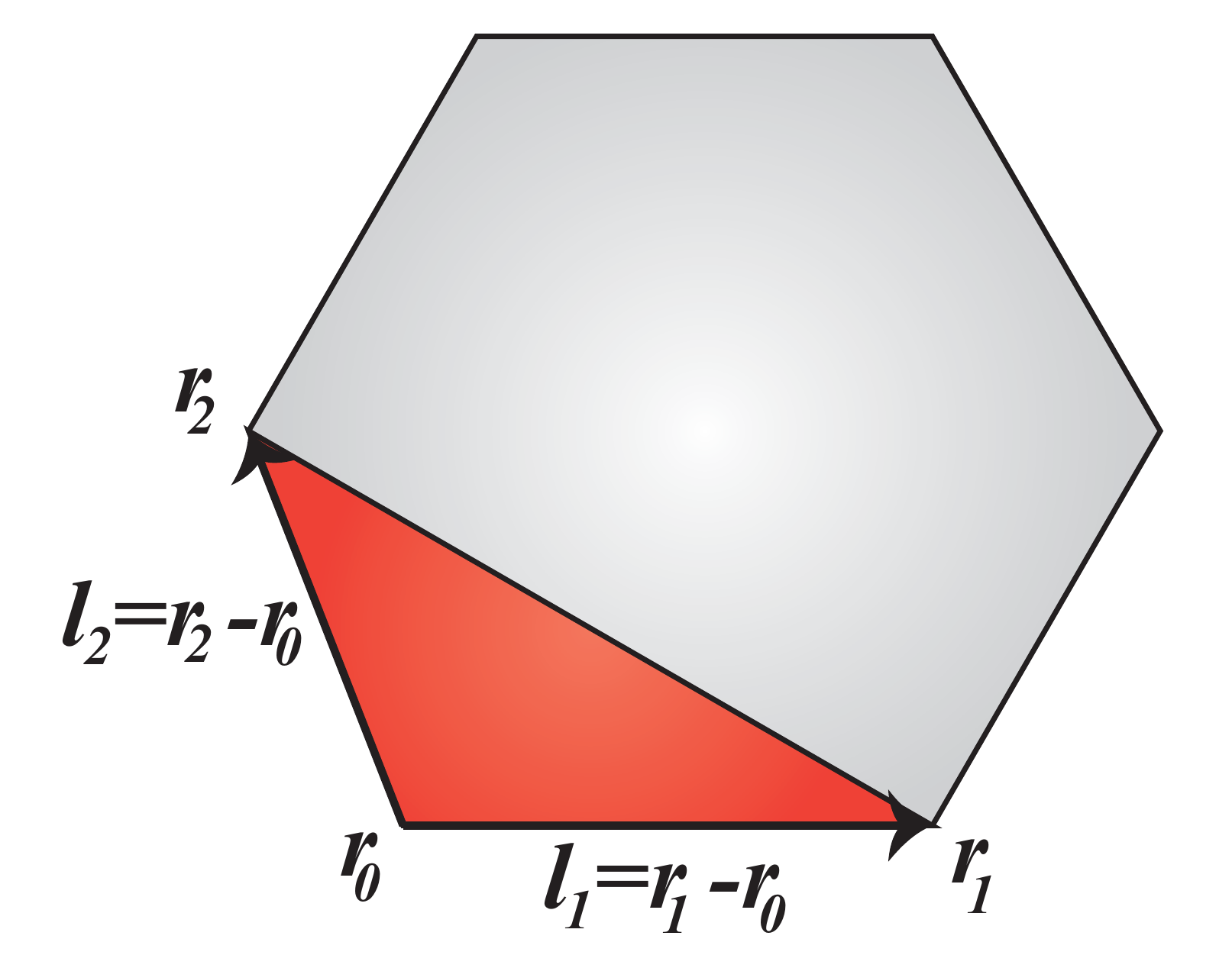}
    \caption{Cartoon of a cell with vertices at positions $\bm{r}_0$, $\bm{r}_1$ and $\bm{r}_2$. Movement of vertex $\bm{r}_0$ affects the lengths of the adjacent edges $\bm{l}_1$ and $\bm{l}_2$ and the area of the shaded triangle bounded by these edges. We assume that the face of the cell is in the $x$-$y$ plane of a standard right-handed coordinate system with the $\hb{z}$ axis projecting out of the plane
. The area of the shaded region is then $\frac{1}{2} \hat{\bm{z}} \cdot \left(\bm{l}_1 \times \bm{l}_2 \right)$.}
    \label{fig:defrl}
\end{figure}

 \subsection{Definition}
Although we will eventually consider generalizations where the forces on vertices cannot be derived from an energy, the vertex model is most commonly stated in terms of an effective energy that is a function of the vertex positions; the final tissue shape is then given by a minimum of the energy function. The form of this energy differs slightly between authors, but a basic version is \cite{Honda:VMenergy}, 
\begin{equation}
	E = \sum_{i}{\Gamma_i l_i} + \frac{K}{2} \sum_{\alpha}{ \left( A_\alpha- A_{0 \alpha}  \right)^2}.
	\label{eq:Energy}
\end{equation}
The first term describes the interfacial tension along the edges and combines the effects of both cell-cell adhesion and actomyosin contractility in the adherens band \cite{Lecuit:actomyocin,Kemkemer:actomyocin}. The sum over $i$ runs over all edges, with edge $i$ having tension $\Gamma_i$ and length $l_i$. The second term describes the energy cost of deforming cells from their preferred area. The sum over $\alpha$ runs over all of the cells in the tissue. $K$ gives the strength of the interaction, $A_{0 \alpha}$ is the preferred area of cell $\alpha$, and $A_\alpha$ is the actual area. A common further simplification is to assume that all edges have the same properties so that $\Gamma_i=\Gamma$ for all $i$; we will call this assumption the \textit{equal tension vertex model}. 

From eq. \ref{eq:Energy} we can immediately find the force on a vertex by taking the derivative with respect to the vertex position, 
\begin{equation}
	\bm{F}_{r_0} = -\frac{\partial E}{\partial \bm{r}_0},
\end{equation}
where $\bm{r}_0$ is the position of the vertex in the two-dimensional plane of the epithelium. We evaluate this force using eq. \ref{eq:Energy}: 
\begin{equation}
	\frac{\partial E}{\partial \bm{r}_0} = K\sum_{[\alpha]}{ \left(A_\alpha-A_{0 \alpha} \right) \frac{\partial A_\alpha}{\partial \bm{r}_0 }}+ \sum_{[i]}{\Gamma _i \frac{\partial l_i}{\partial \bm{r}_0}}.
	\label{eq:dE}
\end{equation}
The movement of a single vertex only effects the lengths and areas of its neighboring cells and edges, so the sums over all edges $i$ and cells $\alpha$ become sums over neighboring edges $[i]$ and cells $[\alpha]$. In order to work out the derivatives it is helpful to introduce some new notation. Let $\bm{l}_i=\bm{r}_i-\bm{r}_0$ be the edge between the vertex at $\bm{r}_0$ and the adjacent vertex at position $\bm{r}_i$, as shown in fig. \ref{fig:defrl}. The cell is taken to be in the two-dimensional $x$-$y$ plane, with $z$ normal to its surface. The change in the edge length is 
\begin{equation}
\frac{\partial l_i}{\partial \bm{r}_0}= -\hb{l}_i,
\end{equation}
where $\hb{l}_i$ is a unit vector which points out from $\bm{r_0}$ along edge $\bm{l}_i$. The only change to the area of the adjacent cells comes from the triangle made by the two edges adjacent to the vertex (shown as the shaded region in fig. \ref{fig:defrl}). The change in the area of this triangle is given by
\begin{align}
	\frac{\partial A}{\partial \bm{r}_0} &= \frac{\partial}{\partial \bm{r}_0} \left[ \frac{1}{2} \hat{\bm{z}} \cdot \left(\bm{l}_1 \times \bm{l}_2 \right) \right] \notag \\
&=   \frac{1}{2} \hb{z} \times ( \bm{l}_2 - \bm{l}_1 ).
	\label{eq:dA}
\end{align}
The change in the energy from a small movement of one vertex is then given by:
\begin{equation}
\frac{\partial E}{\partial \bm{r}_0} = \frac{K}{2} \sum_{[\alpha]}{(A_\alpha-A_{0 \alpha}) \big[\hb{z} \times (\bm{l}_{\alpha 2} - \bm{l}_{ \alpha 1}) \big] + \sum_{[i]}{-\Gamma_i \hb{l}_i}}, \label{eq:dEfinal}
\end{equation}
where $\bm{l}_{\alpha 1}$ and $\bm{l}_{\alpha 2}$ are the two edges which are neighbors of both vertex $\bm{r_0}$ and cell $\alpha$, ordered such that $\hat{\bm{z}}\cdot\left(\bm{l}_{\alpha 1} \times \bm{l}_{\alpha 2} \right) > 0$. Let the pressure in cell $\alpha$  be given by $P_\alpha$. By definition,
\begin{equation}
P_\alpha=-\frac{\partial E}{\partial A_\alpha}=-K(A_\alpha-A_{0 \alpha}). \label{eq:pressure}
\end{equation}
Therefore the total force on any vertex $\bm{r}_0$ is given by
\begin{equation}
	\bm{F}_{r_0} = \sum_{[\alpha]}{\frac{P_\alpha}{2} \big[ \hb{z} \times ( {\bm l_{\alpha2}} - {\bm l_{\alpha1}} ) \big]} + \sum_{[i]}{\Gamma_i \hb{l}_i}.
	\label{eq:VM}
\end{equation}
Note that the direction of the pressure force from a given cell on a given vertex depends on the lengths of the two edges that the cell and vertex share; the force vector does not in general bisect the angle between the two edges. The second term gives the force from the tension on the neighboring edges. 

Although we have derived eq. \ref{eq:VM} from a particular energy function, its physical interpretation, in which each vertex is directly affected by the pressures of the surrounding cells and the tensions of the surrounding edges, suggests a wider validity.  In fact, we can take eq. \ref{eq:VM} to \textit{define} a broader class of vertex models in which the pressure $P_\alpha$ in cell $\alpha$ and the tension $\Gamma_i$ on edge $i$ are given functions of variables that could include edge length and orientation, cell shape, cell types, protein concentrations, and so on.  This class includes as a special case models that posit variants of the energy of eq. \ref{eq:Energy}, like those that include a term quadratic in cell perimeter \cite{Juelicher:VM}; the pressures and tensions are then given as $P_\alpha = -\partial E/\partial A_\alpha$ and $\Gamma_i = \partial E/\partial l_i$.  A vertex model defined directly in terms of the force on a vertex, however, also encompasses models that cannot be derived from any underlying global energy, including examples in which tensions depend on the protein concentration on an edge \cite{Lubensky:2012} or on edge orientation.  In the remainder of this section and in section \ref{sec:stab}, we will formulate vertex dynamics and the conditions for local stability of fourfold vertices in terms of arbitrary pressures $P_\alpha$ and tensions $\Gamma_i$; starting in section \ref{sec:plateau}, we will then turn to consider what our stability conditions imply for some specific, simple choices of the $\Gamma_i$.

\subsection{Dynamics}
\begin{figure*}[!ht]
		\centering
    \includegraphics[width=0.9\textwidth]{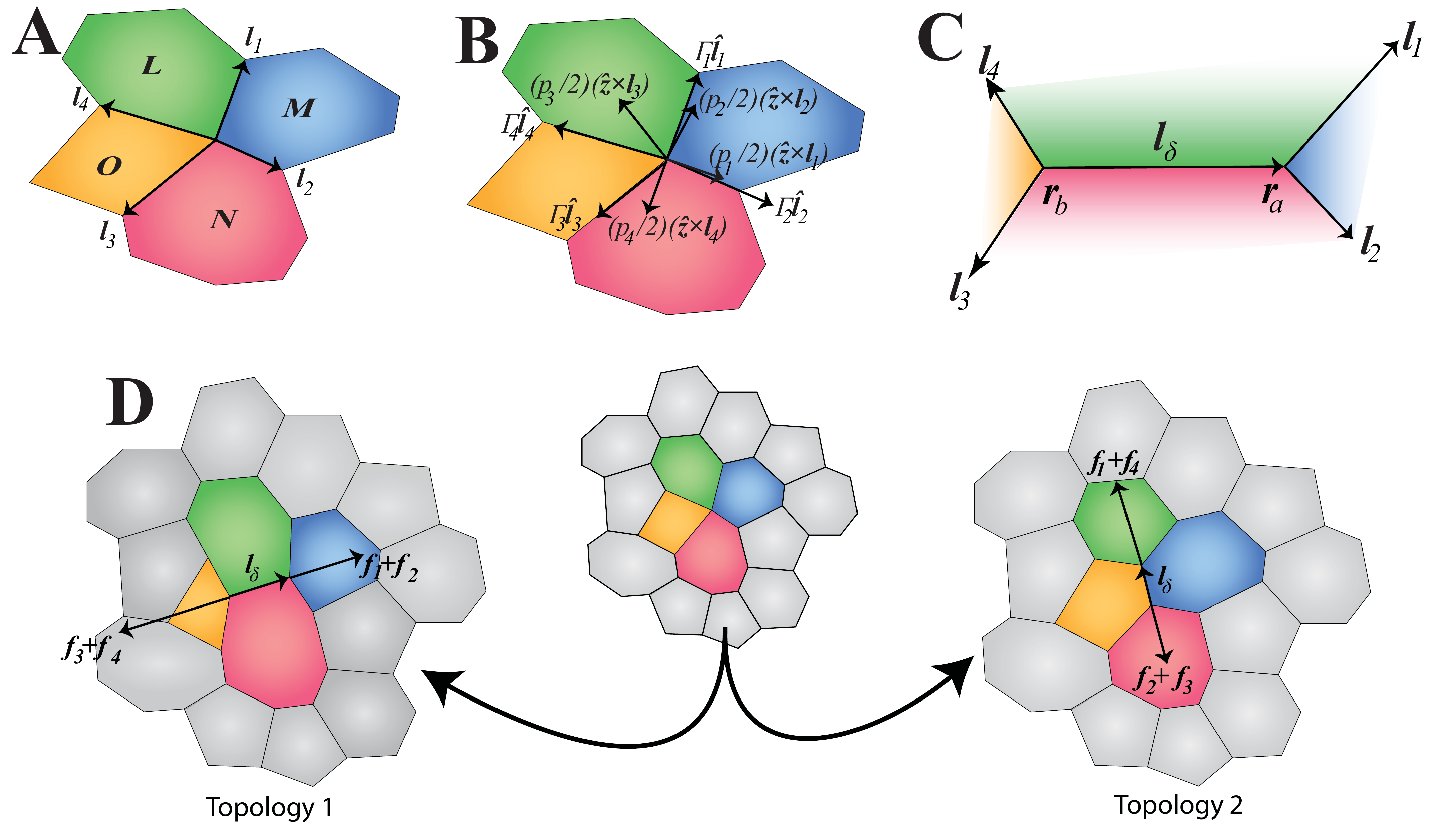}
    \caption{ 
		$\bf{A:}$ Cartoon of a fourfold vertex with neighboring cells $L$, $M$, $N$, $O$ and edges $\bm{l}_1$, $\bm{l}_2$, $\bm{l}_3$, $\bm{ l}_4$. Note that the direction of the edges is outward from the vertex, and the edges are numbered clockwise. \newline $\bf{B:}$ Cartoon showing eight different forces acting on the fourfold vertex, two associated with each edge (eq. \ref{eq:force}). The four edges produce a tension $\Gamma_i\hb{l}_i$. The effect of the pressures from the four cells can be written in terms of the pressure differences across the edges; if $p_i$ is the pressure difference across edge $i$, we can view the pressures as exerting a force $\frac{p_i}{2}(\hb{z}\times\bm{l}_i)$ perpendicular to each edge. \newline $\bf{C:}$ Cartoon of two threefold vertices which share an edge $\bm{l}_\delta$. As its length $l_\delta$ shrinks to zero, the vertices $\bm{r}_a$ and $\bm{r}_b$ will merge to form a single fourfold vertex. \newline $\bf{D:}$ Cartoon of the resolution of a fourfold vertex. The fourfold vertex (center) can break apart into two threefold vertices in either of two topologies (left, right).  In each case, we can associate a told force $\bm{f}_i$ with each edge $i$ that includes tension and pressure jump contributions.  In Topology 1 (left), forces $\bm{f}_3$ and $\bm{f}_4$ act one of the new vertices and forces $\bm{f}_1$ and $\bm{f}_2$ act on the other new vertex, so that the net force trying to extend the new edge $\bm{l}_\delta$ is $\bm{f}_1+\bm{f}_2-\bm{f}_3-\bm{f}_4$; this is counteracted by the tension $\Gamma_\delta$ on the new edge (eq.~\ref{eq:deltadot}).  The situation is the same in Topology 2 (right), but with the edges paired differently.
}
    \label{fig:stability}
\end{figure*}

To determine the motion of the vertices we make the common assumption that the vertices experience a drag force proportional to their velocity, so that 
\begin{equation}
\bm{F}_{r_0}=\mu \dot{\bm{r}_0}
\label{eq:local-dissip}
\end{equation} 
where $\mu$ is the drag constant. (Other assumptions about the form of dissipation have also been proposed \cite{Kawasaki:89} but will not be considered here.) An arbitrary vertex then moves according to
\begin{equation}
\dot{\bm{r}}_0 = \frac{1}{\mu} \left(  \sum_{[\alpha]}  {\frac{P_\alpha}{2} \big[ \hb{z} \times ( {\bm l_{\alpha2}} - {\bm l_{\alpha1}} ) \big]}  + \sum_{[i]}  {\Gamma_i \hb{l}_i}   \right).
\end{equation}
 The sum over the neighboring cells $[\alpha]$ includes taking the difference between neighboring edges, which also appear in the second sum over the neighboring edges $[i]$. By expressing the forces in terms of the pressure difference across an edge, we can combine these sums into a single sum over neighboring edges. To illustrate how the sums are merged let us consider an arbitrary vertex, which happens to be fourfold, with cells L, M, N, O, and edges 1, 2, 3, 4 as shown in fig. \ref{fig:stability}A. Explicitly writing out the force on the vertex from eq. \ref{eq:VM} gives
\begin{align}
	\bm{F}_{r_0}&= \Gamma_1 \hb{l}_1+\Gamma_2 \hb{l}_2+ \Gamma_3\hb{l}_3 +\Gamma_4 \hb{l}_4 \notag \\
	&+\frac{P_L}{2} \hb{z} \times (\bm{l}_4-\bm{l}_1) + \frac{P_M}{2} \hb{z} \times (\bm{l}_1-\bm{l}_2) \notag \\
	&+ \frac{P_N}{2} \hb{z} \times (\bm{l}_2-\bm{l}_3) + \frac{P_O}{2}\hb{z}\times (\bm{l}_3-\bm{l}_4).
\end{align}
We regroup the terms so that each term contains only one edge. 
\begin{align}
	\bm{F}_{r_0} &= \Gamma_1 \hb{l}_1+\Gamma_2 \hb{l}_2+ \Gamma_3\hb{l}_3 +\Gamma_4 \hb{l}_4 \notag \\
	&+ \frac{P_L-P_O}{2}(\hb{z} \times \bm{l}_4) +\frac{P_M-P_L}{2}(\hb{z}\times \bm{l}_1) \notag \\
	&+\frac{P_N-P_M}{2}(\hb{z} \times \bm{ l}_2) + \frac{P_O-P_N}{2}(\hb{z} \times \bm{ l}_3)  
\end{align}
We can further simplify this expression by introducing the notation
\begin{equation}
p_i=P_\alpha - P_{\alpha'},
\end{equation}
where $\alpha$ and $\alpha'$ are the cells on either side of edge $i$, so that $p_i$ represents the difference in pressure across an edge taken counterclockwise around the vertex. For example in the configuration show in fig. \ref{fig:stability}B, $p_1=P_M-P_L$. In this simplified notation the force on our fourfold vertex is
\begin{align}
\bm{F}_{r_0}&=\Gamma_1\hb{l}_1 + \frac{p_1}{2}(\hb{z} \times \bm{l}_1)+ \Gamma_2\hb{l}_2 + \frac{p_2}{2}(\hb{z} \times \bm{l}_2) \notag \\
&+ \Gamma_3\hb{l}_3+ \frac{p_3}{2}(\hb{z} \times \bm{l}_3)+ \Gamma_4\hb{l}_4 + \frac{p_4}{2}(\hb{z} \times \bm{l}_4).
\end{align}
In general we can write the force on any vertex $r_0$ as
\begin{equation}
\bm{F}_{r_0} = \sum_{[i]} \left[ {\Gamma_i \hb{l}_i} +\frac{p_i}{2}(\hb{z} \times \bm{l}_i) \right]. \label{eq:force}
\end{equation}

\section{Fourfold vertex stability}
\label{sec:stab}
In this section we will work out the criteria which a fourfold vertex must satisfy in order to be stable. As preparation, we first in section \ref{sec:stab-dyn} examine the dynamics of neighboring threefold vertices as the length of their shared edge approaches zero.  When the edge length reaches zero, a fourfold vertex can be formed; once formed, it can either persist as a fourfold vertex, or it can resolve into threefold vertices in one of two possible topologies (fig. \ref{fig:stability}D). We call a fourfold vertex stable if, when it is broken apart into two threefold vertices separated by a small shared edge $\bm{l}_\delta$, the forces on the two threefold vertices push them back together, causing the edge $\bm{l}_\delta$ to shrink to zero; this condition must hold for \textit{both} possible resolution topologies. Section \ref{sec:stab-defn} makes this notion of stability more precise and addresses some technical questions that it raises.  Finally, in section \ref{sec:stab-conditions} we work out the criterion for a fourfold vertex to be stable against resolving in one topology.  The criterion for the other topology then follows immediately, and combining the two gives us our final set of stability conditions.

\subsection{Dynamics of a small edge}
\label{sec:stab-dyn}
Consider the dynamics of a pair of threefold vertices $\bm{r}_a$ and $\bm{r}_b$ which share an edge as shown in fig. \ref{fig:stability}C. Define the shared edge $\bm{l}_\delta$ as $\bm{l}_\delta = \bm{r}_a-\bm{r}_b$. This edge evolves according to
\begin{equation}
\dot{\bm{l}}_\delta=\dot{\bm{r}}_a-\dot{\bm{r}}_b = \frac{1}{\mu} \left(\bm{F}_{r_a} - \bm{F}_{r_b} \right),
\end{equation}
where $\bm{F}_{r_a}$ and $\bm{F}_{r_b}$ are the forces on the two vertices given by eq. \ref{eq:force}. This equation uses the conventions that the direction of $\bm{l}_i$ is taken outward from the vertex and that the pressures $p_i$ are taken counterclockwise around the vertex.  As we are now dealing with two vertices, we modify these conventions slightly to (arbitrarily) take vertex $a$ as the reference vertex, so that the contribution from the tension on $\bm{l}_\delta$ is positive in the force on vertex $a$ and negative in the force on vertex $b$.  Similarly, we define the pressure difference $p_\delta$ across $\bm{l}_\delta$ to be taken counterclockwise around $\bm{r}_a$; because both $p_\delta$ and $\bm{l}_\delta$ then flip signs, the pressure difference across the shared edge contributes with the same sign to the forces on both vertices.  Substituting,
\begin{align}
	\dot{\bm{l}}_\delta &= \frac{1}{\mu} \bigg[ \Gamma_1 \hb{l}_1 + \frac{p_1}{2}(\hb{z} \times \bm{l}_1) \notag \\
	&+\Gamma_2\hb{l}_2+ \frac{p_2}{2}(\hb{z} \times \bm{l}_2) -\Gamma_{\delta}\hb{l}_\delta + \frac{p_\delta}{2}(\hb{z} \times \bm{l}_\delta) \bigg] \notag \\
	&-\frac{1}{\mu} \bigg[ \Gamma_3 \hb{l}_3 + \frac{p_3}{2}(\hb{z} \times \bm{l}_3) \notag \\
	&+\Gamma_4\hb{l}_4+ \frac{p_4}{2}(\hb{z} \times \bm{l}_4) + \Gamma_{\delta}\hb{l}_\delta + \frac{p_\delta}{2}(\hb{z} \times \bm{l}_\delta) \bigg] \; ,
\end{align}
where $\Gamma_{\delta}$ is the tension of the shared edge. 

Define $\bm{f}_i$ as the contribution to the force associated with edge $i$, $\bm{f}_i=\Gamma_i\hb{l}_i + \frac{p_i}{2}(\hb{z} \times \bm{l}_i)$.  Then the shared edge follows the equation of motion
\begin{equation}
	\mu \dot{\bm{l}}_\delta= \bm{f}_1+\bm{f}_2-\bm{f}_3-\bm{f}_4- 2\Gamma_\delta \hb{l}_\delta.
	\label{eq:deltadot}
\end{equation}  
The forces $\bm{f}_1$ through $\bm{f}_4$ can in general depend on $\bm{l}_\delta$, and indeed on the positions of all the other vertices.  Importantly, however, all four forces generically approach a finite, nonzero limit as $l_\delta \rightarrow 0$.  (This contrasts with the situation in a standard linear stability problem in which forces would go to zero linearly with $l_\delta$.)    As we discuss in more detail in the next section, when looking at vertex stability we will always be interested in the limit of small $l_\delta$.  To leading order in this limit, $\bm{f}_1$ through $\bm{f}_4$ can thus be evaluated at $\bm{l}_\delta = 0$ and treated as constants.  Our subsequent development always assumes that this limit has been taken.

\subsection{Defining fourfold vertex stability}
\label{sec:stab-defn}

To think about vertex stability, we would like to imagine, informally, that the fourfold vertex is constantly subject to noise or other small perturbations and that, from time to time, these perturbations cause it to break up into a pair of barely separated threefold vertices, with more or less random topology and orientation.  If, for small enough perturbations, the fourfold vertex always re-forms, then we should call it stable.  On the other hand, if the vertex dynamics ever tend to move the two newly formed threefold vertices apart, we would like to call the fourfold vertex unstable.  Thus, to define stability more carefully, we ask what happens if, at some instant, a fourfold vertex is replaced by two threefold vertices whose separation $\bm{l}_\delta$ is infinitesmally small (and whose average position is at most infinitesmally different from the position of the original fourfold vertex).  The separation $\bm{l}_\delta$ is then allowed to evolve according to eq.~\ref{eq:deltadot}.  If, when the magnitude $l_\delta = |\bm{l}_\delta|$ of the separation between the two vertices is small enough, its time derivative $\mathrm{d}l_\delta/\mathrm{d} t$ is always negative, for both possible resolution topologies and for any choice of orientation $\hb{l}_\delta$, then the fourfold vertex is stable.  If there is any choice of separation orientation $\hb{l}_\delta$ and topology for which $\mathrm{d}l_\delta/\mathrm{d} t$ remains positive for arbitrarily small $l_\delta$, then the vertex is unstable.  Finally, if, as $l_\delta$ goes to zero, $\mathrm{d}l_\delta/\mathrm{d} t$ approaches zero for one or more choices of $\hb{l}_\delta$ but is otherwise negative, then the fourfold vertex is either marginally stable or marginally unstable, and the calculation must be pursued to higher order in $l_\delta$ than we consider in this paper.

Several aspects of this definition of stability deserve further comment.  First, the positions of vertices, and hence the forces in eq.~\ref{eq:deltadot} and the stability of a given fourfold vertex, may change with time.  Thus, vertex stability is an instantaneous notion, and we should really talk about the stability or instability of a vertex at some time $t_0$; in particular, in the most general case it is possible for a fourfold vertex to be stable at time $t_0$ but then, because of the natural time evolution of the cell packing and without any change in system parameters, to go unstable at some later time $t_1 > t_0$.  Second, to determine stability at some time $t_0$ in practice, we evaluate the forces in eq.~\ref{eq:deltadot} as if both new vertices were located exactly at the position of the fourfold vertex in question and all other vertices were frozen at their positions at time $t_0$.  This is appropriate because of the observation that all of the force terms in eq.~\ref{eq:deltadot} generically have finite, nonzero limits as $l_\delta$ approaches zero at fixed orientation $\hb{l}_\delta$.  Except in the marginal case described in the preceding paragraph, for small enough $l_\delta$ these finite terms must dominate any corrections due to infinitesmal deviations of vertices from their positions at time $t_0$.  Finally, the same reasoning explains why we can focus exclusively on the dynamics of the separation $\bm{l}_\delta$ and can ignore the possibility of collective instability modes that involve the motion of many vertices:  As long as $\mathrm{d}l_\delta/\mathrm{d} t$ is finite and nonzero as $l_\delta \rightarrow 0$, infinitesmal perturbations to other vertex positions can change its magnitude infinitesmally, but cannot affect its sign.

\subsection{Stability conditions}
\label{sec:stab-conditions}
In accordance with the notion of stability described in the previous section, we now imagine that the fourfold vertex momentary splits into two infinitesimally close threefold vertices as shown in fig. \ref{fig:stability}D. In order for the vertex to be stable we want the vertex dynamics to force the two vertices back together. We know that in general the vertices' shared edge evolves according to eq. \ref{eq:deltadot}. Let $\bm{\mathcal{F}}=\bm{f}_1+\bm{f}_2-\bm{f}_3-\bm{f}_4$, and let $\theta$ be the angle between $\bm{\mathcal{F}}$ and the edge $\bm{l}_\delta$. The time derivatives of the length $l_\delta$ and direction $\theta$ of the newly formed edge are given by
\begin{align}
l_\delta {\dot \theta} &=  -\frac{\mathcal{F}}{\mu} \sin{\theta} \label{eq:theta}
\\ \dot{l}_ \delta &=  \frac{\mathcal{F}}{\mu} \cos{\theta} - \frac{2 \Gamma_{\delta}}{\mu}, \label{eq:delta},
\end{align}
where $\mathcal{F} = |\bm{\mathcal{F}}|$.  From eq. \ref{eq:delta}, we conclude that the shared edge grows the fastest when $\theta=0$. Therefore it is sufficient to look at new edges which form along the line of force $\bm{\mathcal{F}}$ to prove stability. In this case eq. \ref{eq:delta} reduces to
\begin{equation}
	\dot{l}_\delta=\frac{\mathcal{F}-2\Gamma_{\delta}}{\mu}.
\end{equation}
The edge shrinks whenever 
\begin{equation}
\Gamma_{\delta} > \frac{\mathcal{F}}{2}.
\label{eq:stability}
\end{equation}

We note in passing that when the forces are derived from an energy, we can also see that eq. \ref{eq:stability} must be the stability criterion by looking at the energy. The change in energy to lowest order in $\bm{l}_\delta$ is $\delta E=\bm{\mathcal{F}}\cdot\bm{l}_\delta-2\Gamma_\delta |\bm{l}_\delta|$, so that the change in energy is negative whenever $\Gamma_\delta > \frac{\bm{\mathcal{F}}\cdot\bm{l}_\delta}{2l_\delta}$. The right hand side is maximized when $\bm{\mathcal{F}}$ is in the direction of the new edge, and so the vertex is stable whenever $\Gamma_\delta>\frac{\mathcal{F}}{2}$.

It is important to remember that the vertex can resolve in two different topologies (fig. \ref{fig:stability}D) which have different $\bm{\mathcal{F}}$. Therefore, for the vertex to be stable, both of the following conditions must be met: 
\begin{align}
	\Gamma_{\delta} &\geq \frac{|\bm{f}_1+\bm{f}_2-\bm{f}_3-\bm{f}_4|}{2} \label{eq:c1} \\
	\Gamma_{\delta} &\geq \frac{|\bm{f}_2+\bm{f}_3-\bm{f}_1-\bm{f}_4|}{2} \label{eq:c2}. 
\end{align}
The condition in eq. \ref{eq:c1} ensures that the fourfold vertex is stable against resolution into two threefold vertices in topology 1 (fig. \ref{fig:stability}D), by enforcing the stability criterion derived in eq. \ref{eq:stability}. Similarly, condition \ref{eq:c2} ensures that the vertex is stable against resolution in topology 2 (fig. \ref{fig:stability}D).

Except in section \ref{sec:nonequilibr}, we will primarily be interested in what follows in the stability of fourfold vertices that are in mechanical equilibrium---that is, on which the net force is zero.  (Because of our assumption of local dissipation at the vertex, eq.~\ref{eq:local-dissip}, mechanical equilibrium of a vertex is equivalent to its being stationary.)  If this additional condition holds, then $\bm{f}_1+\bm{f}_2 + \bm{f}_3 + \bm{f}_4 = \bm{0}$, and one can replace $-\bm{f}_3 - \bm{f}_4$ by $\bm{f}_1+\bm{f}_2$ and $-\bm{f}_1 - \bm{f}_4$ by $\bm{f}_2+\bm{f}_3$ in eqs. \ref{eq:c1}--\ref{eq:c2} (thereby removing all dependence on $\bm{f}_4$ in both inequalities).  The two inequalities can then be rewritten explicitly in terms of the $p_i$ and $\Gamma_i$ as 
\begin{align}
	\Gamma_\delta & \geq |\Gamma_1 \hb{l}_1 + \frac{p_1 l_1}{2}(\hb{z} \times \hb{l}_1)+ \Gamma_2 \hb{l}_2 + \frac{p_2 l_2}{2}(\hb{z} \times \hb{l}_2)| \label{eq:c1general} \\ 
	\Gamma_\delta & \geq |\Gamma_3 \hb{l}_3 + \frac{p_3 l_3}{2}(\hb{z} \times \hb{l}_3)+ \Gamma_2 \hb{l}_2 + \frac{p_2 l_2}{2}(\hb{z} \times \hb{l}_2)| \label{eq:c2general}.
\end{align}
Similarly, the equation of mechanical equilibrium takes the form
\begin{align}
	0 & = \Gamma_1 \hb{l}_1 + \frac{p_1 l_1}{2}(\hb{z} \times \hb{l}_1)+ \Gamma_2 \hb{l}_2 + \frac{p_2 l_2}{2}(\hb{z} \times \hb{l}_2) \notag \\
	& + \Gamma_3 \hb{l}_3 + \frac{p_3 l_3}{2}(\hb{z} \times \hb{l}_3)+ \Gamma_4 \hb{l}_4 + \frac{p_4 l_4}{2}(\hb{z} \times \hb{l}_4)\label{eq:c3general}.
\end{align}
A physical interpretation of these stability conditions is that eq. \ref{eq:c1general} and eq. \ref{eq:c2general} require that the tension on the new edge is high enough that it counteracts the forces from the tensions and pressure differences across the edges. This stops the vertex from resolving in either of the possible topologies. Equation \ref{eq:c3general} constrains the vertex to be in mechanical equilibrium.

\section{No stable, stationary fourfold vertices exist in Plateau's model}
\label{sec:plateau}
\begin{figure}[]
		\centering
    \includegraphics[width=0.35\textwidth]{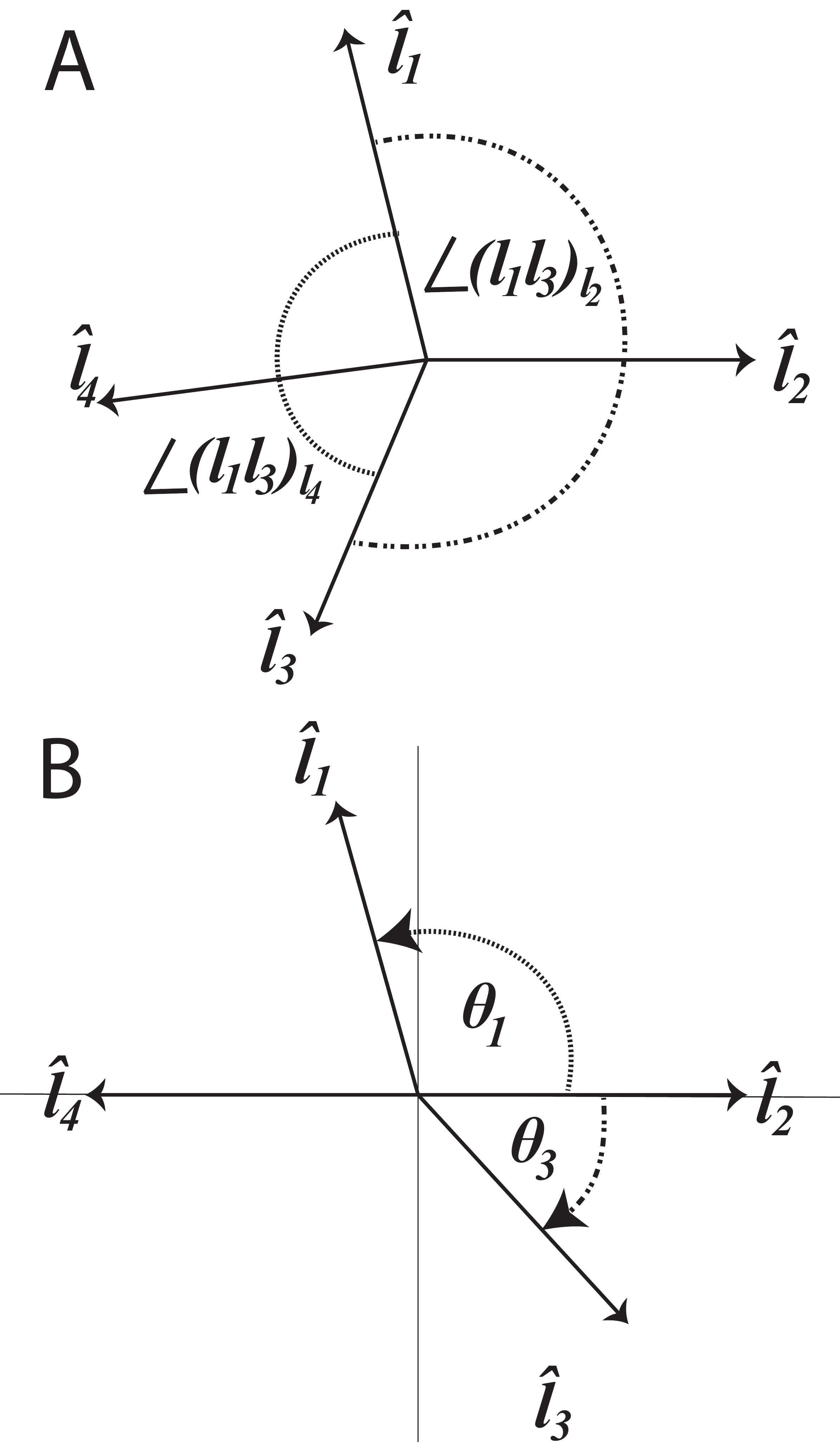}
    \caption{ $\bf{A:}$ The two angles $\angle (l_1 l_3)_{l_2}$ and $\angle (l_1 l_3)_{l_4}$ between the non-adjacent edges $\bm{l}_1$ and $\bm{l}_3$ are shown. The quantities $\angle(l_i l_j)_{l_k}$ are defined as the unsigned magnitudes of the angles, so $\angle(l_i l_j)_{l_k}=\angle(l_j l_i)_{l_k}$. The angles $\angle (l_1 l_3)_{l_2}$ and $\angle (l_1 l_3)_{l_4}$ together make a full circle, implying $\angle (l_1 l_3)_{l_2}+\angle (l_1 l_3)_{l_4}=2\pi$. $\bf{B:}$ The angles $\theta_1$ and $\theta_3$ are defined in the usual manner as the signed angles between the positive $x$ axis (which here coincides with $\bm{l}_2$) and, respectively, $\bm{l}_1$ and $\bm{l}_3$.  Hence, as drawn, $\theta_1 > 0$ and $\theta_3 < 0$.
		}
    \label{fig:plateau}
\end{figure}

In section \ref{sec:proof} we will show that there are no stable, stationary fourfold states under the condition that all of the edges have the same tension, and in section \ref{sec:stabelex} we will give some examples of stable fourfold vertices that arise when we lift this requirement. In this section, we first work through the simplest special case of eqs. \ref{eq:c1general}-\ref{eq:c3general} to give the reader some intuition about the main proof presented in section \ref{sec:proof}. We will use the same structure for our proof in both sections. 

The simplest possible situation is one in which $p_i=0$ and $\Gamma_i=\Gamma$. This is equivalent to Plateau's model of a dry foam, because in Plateau's model the pressure does not affect the motion of the vertices directly but instead changes the angle of vertices' neighboring edges. (Equivalently, changes to cell areas when a new edge is created are higher order in $\delta$ than are changes in edge lengths and thus can be neglected in calculations of vertex stability in Plateau's model~\cite{Hutzler:Proof}.) 

To begin the proof that stationary vertices cannot be stable in this case we first rewrite the criteria for stability from eqs. \ref{eq:c1general}-\ref{eq:c3general}.  After dividing by $\Gamma$, we have
\begin{align}
	1 & \geq |\hb{l}_1 + \hb{l}_2| \label{eq:cond1plat} \\ 
	1 & \geq |\hb{l}_3 + \hb{l}_2| \label{eq:cond2plat} \\
	0 &= \hb{l}_1 + \hb{l}_2+\hb{l}_3 +\hb{l}_4. \label{eq:cond3plat}
\end{align}
As in fig. \ref{fig:stability}A, we label the edges in the clockwise direction from 1 through 4, and we assume that each pair of successively numbered edges bounds a single cell: cell $M$ lies between $\bm{l}_1$ and $\bm{l}_2$, cell $N$ lies between $\bm{l}_2$ and $\bm{l}_3$, and so on.  For our model to be physically reasonable we cannot have two or more cells occupying the same space, so we must reject any configurations in which edge 1 moves through edge 2 in such a way that cell $M$ inverts and partially overlaps cell $N$. In order to avoid such unphysical overlap, we require that the ordering of the edges around the vertex remain fixed, and thus in particular that the labels 1 through 4 always appear in increasing order in the clockwise direction.


As shown in fig. \ref{fig:plateau}A, the non-adjacent edges 1 and 3 are separated by two angles, one encompassing edge 2 and the other encompassing edge 4, which together make up a full circle.  We call the (necessarily positive) magnitudes of these two angles $\angle(l_1 l_3)_{l_2}$ and $\angle(l_1 l_3)_{l_4}$; more generally, we refer to the magnitude of the angle between non-adjacent edges $\bm{l}_i$ and $\bm{l}_j$ that encompasses $\bm{l}_k$ as $\angle(l_i l_j)_{l_k}$.

\begin{figure}[]
		\centering
    \includegraphics[width=0.45\textwidth]{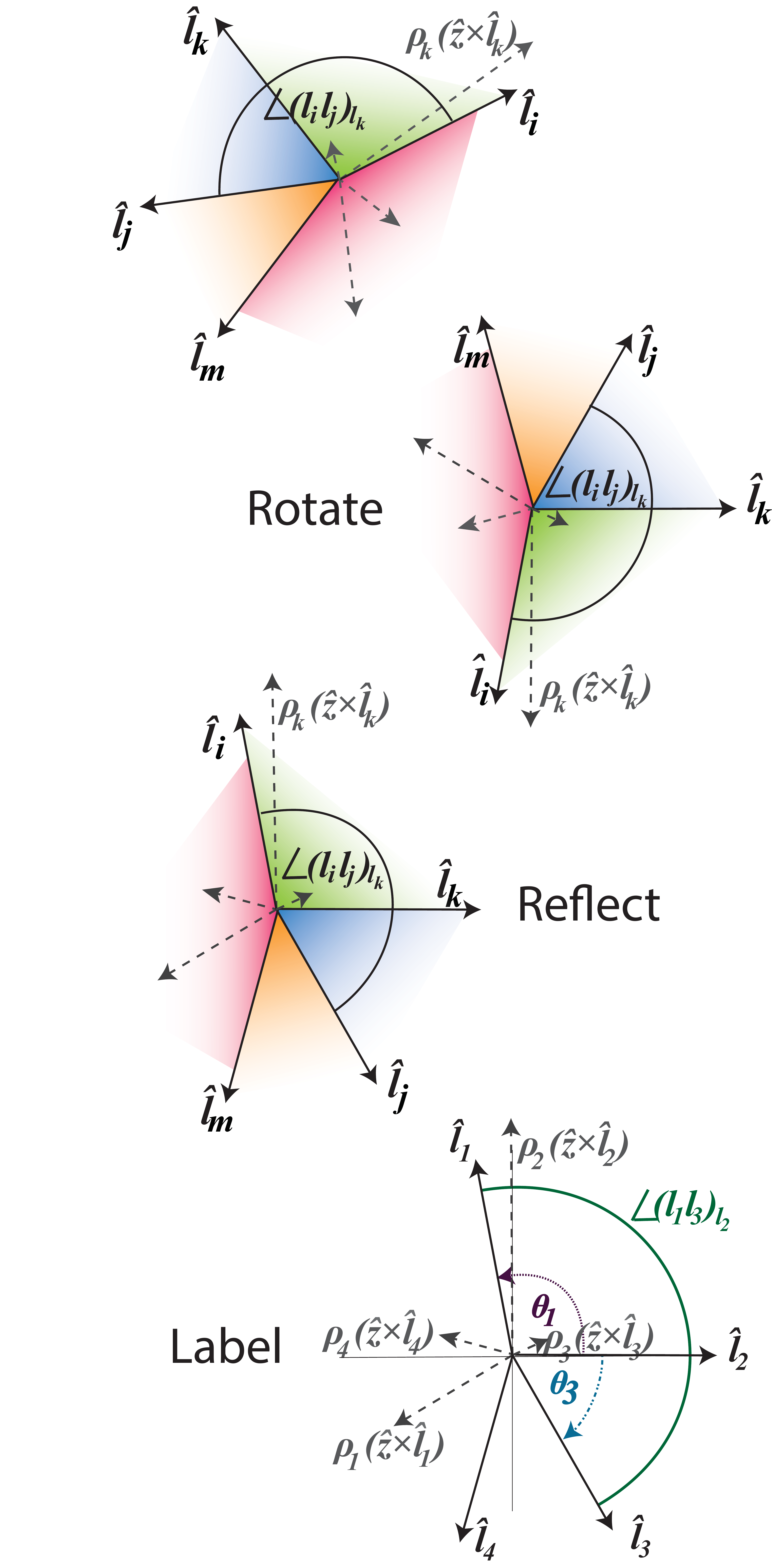}
    \caption{ 
		Cartoon of the procedure in section \ref{proof:polar} to exploit the symmetries of the problem in order to reduce the number of free variables. An arbitrary angle $\angle(l_il_j)_{l_k}$ between non adjacent edges can be transformed under rotations and reflections such that $\bm{l}_k$ lies on the positive x-axis, and $\rho_k(\hb{z}\times\hb{l}_k)$ lies on the positive y-axis. The edges can then be renumbered from 1 to 4 in the clockwise direction. $\theta_1$ and $\theta_3$ are the signed angles that $\hb{l}_1$ and $\hb{l}_3$ make with the positive $x$ axis, as shown, so that the magnitude of the angle between $\hb{l}_1$ and $\hb{l}_3$ is $\angle(l_1l_3)_{l_2}=\theta_1-\theta_3$.  
		}
    \label{fig:symmetry}
\end{figure}

To show that a fourfold vertex cannot be stable in the Plateau model, begin by taking an arbitrary pair of non-adjacent edges $\bm{l}_i$ and $\bm{l}_j$. Either $\angle(l_i l_j)_{l_k}\leq\pi$ or $\angle(l_i l_j)_{l_m}\leq\pi$ (where $\bm{l}_k$ and $\bm{l}_m$ are the other two edges at the vertex); we choose without loss of generality to label the edges so that $\angle(l_i l_j)_{l_k}\leq\pi$. We may then apply a rotation followed by (if needed) a reflection to the fourfold vertex and relabel the edges so that $\angle(l_i l_j)_{l_k}$ becomes $\angle(l_1 l_3)_{l_2}$ and $\hb{l}_2 = \hb{x}$ (fig. \ref{fig:symmetry} and section \ref{proof:polar}).  Let $\theta_i$ be the signed angle between edge $i$ and the $x$ axis, as shown in fig \ref{fig:plateau}B.  Then $\theta_1>0$ and $\theta_3<0$. (Note also that because $\angle(l_1 l_3)_{l_2} \leq \pi$ by assumption, neither $\theta_1$ nor $\theta_3$ can have magnitude larger than $\pi$.)  We will continue to use this convention in section \ref{sec:proof}. 

The next step is to convert eqs. \ref{eq:cond1plat} and \ref{eq:cond2plat} to polar coordinates,
\begin{align}
	1 & \geq \left(\cos{\theta_1}+1 \right)^2+\left (\sin{\theta_1} \right )^2 \\ 
	1 & \geq \left(\cos{\theta_3}+1 \right)^2+\left (\sin{\theta_3} \right )^2,
\end{align}
and to solve the system of inequalities.  In this case we can immediately deduce that, for these conditions to hold and the vertex to be stable, we must have $\theta_1 \geq 2 \pi/3$ and $\theta_3 \leq -2 \pi/3$. It follows that $\angle(l_1 l_3)_{l_2} = \theta_1 - \theta_3 \geq 4 \pi/3$, which contradicts our initial assumption that $\angle(l_1l_3)_{l_2} \leq \pi$.  Hence, the vertex must be unstable. 

Our proof that there are no stable states in the equal tension vertex model will follow the same basic structure. First we will express the conditions \ref{eq:c1general}-\ref{eq:c2general} in polar coordinates. We will solve the resulting system of inequalities to get bounds on the angle between any two non-adjacent edges $\angle (l_i l_j )_{l_k}$. We will then show that the given bounds lead to a contradiction.

\section{No stable, stationary fourfold vertices exist in the equal tension vertex model}
\label{sec:proof}

Throughout section \ref{sec:proof} we will work with a special case of the vertex model, which we call the equal tension vertex model, which shares important features with the Plateau model of foams. In the equal tension vertex model, as in the Plateau model, every edge has the same tension $\Gamma_i=\Gamma$; unlike the Plateau model, however, the equal tension vertex model does not put any restrictions on the cell pressures $P_\alpha$.  In this section, we consider only fourfold vertices that are stationary and in mechanical equilibrium.

Our argument that such fourfold vertices can never be stable in the equal tension model proceeds as follows: In section \ref{proof:note} we introduce the variables $\rho_i$, which are dimensionless ratios of an edge's length, tension, and pressure difference. This reduces the number of variables in the problem to eight (four edge directions and four $\rho_i$). In section \ref{proof:polar}, we express the stability conditions \ref{eq:c1general}-\ref{eq:c2general} in polar coordinates and use the symmetries of the problem to reduce the number of free variables to seven. In section \ref{proof:calc} we analyze the resulting system of inequalities, concluding that fourfold vertices are unstable unless $\angle (l_i l_j)_{l_k} = \pi$ for any choices of non-adjacent edges $\bm{l}_i$ and $\bm{l}_j$ and intervening edge $\bm{l}_k$. Finally, in section \ref{sec:contradiction} we show that if $\angle (l_i l_j)_{l_k} = \pi$ for all pairs of non-adjacent edges, it is impossible to satisfy all three stability and equilibrium conditions \ref{eq:c1general}-\ref{eq:c3general}.  Thus, no stable, stationary fourfold vertices are possible in the equal tension model. 

\subsection{Streamlining notation}
\label{proof:note}

We begin by writing a more compact version of the general stability conditions given in eqs. \ref{eq:c1general} - \ref{eq:c2general}. Let
\begin{equation} 
\rho_i=\frac{p_i |\bm{l}_i|}{2 \Gamma},
\end{equation}
be a scalar which is proportional to the force exerted by the pressure difference across edge $i$. Recall that the pressure difference is taken counterclockwise around vertex $\bm{r}_a$ (fig.~\ref{fig:stability}C), so the sign of $\rho$ depends on which neighboring cell has the higher pressure. The stability criteria can then be expressed as 
\begin{align}
	1 & \geq |\hb{l}_1 + \rho_1(\hb{z} \times \hb{l}_1)+ \hb{l}_2 + \rho_2(\hb{z} \times \hb{l}_2)| \label{eq:cond1} \\ 
	1 & \geq |\hb{l}_3 + \rho_3(\hb{z} \times \hb{l}_3)+ \hb{l}_2 + \rho_2(\hb{z} \times \hb{l}_2)| \label{eq:cond2} \\
	0 &= \hb{l}_1 + \rho_1(\hb{z} \times \hb{l}_1)+ \hb{l}_2 + \rho_2(\hb{z} \times \hb{l}_2) \notag \\
	  &+ \hb{l}_3 + \rho_3(\hb{z} \times \hb{l}_3)+ \hb{l}_4 + \rho_4(\hb{z} \times \hb{l}_4)\label{eq:cond3}.
\end{align}
By absorbing the lengths of the edges into the coefficients $\rho_i$, the problem is now poised entirely in terms of unit vectors. The problem is reduced to eight variables: the four angles of the edges with respect to the x-axis $\theta_1, \theta_2, \theta_3, \theta_4$, and the four $\rho$ coefficients.

As an aside, if we further assume the pressures have the simple form of eq.~\ref{eq:pressure} and express the $\rho_i$ in terms of the areas of the cells,
\begin{align}
	\rho_i &=\frac{p_i l_i}{2 \Gamma} \notag \\
	&=\frac{\left(P_\alpha-P_{\alpha'} \right) l_i}{2 \Gamma} \notag \\
	&=\frac{K l_i \left[ (A_{\alpha'}-A_{\alpha}) + (A_{0 \alpha}-A_{0\alpha'})\right]}{2 \Gamma}, \label{eq:rhodef}
\end{align}
it becomes clear that the preferred area $A_{0\alpha}$ of the cells does not affect the stability in the common case in which $A_{0\alpha}$ is the same for all cells. 

\subsection{Exploiting symmetries}
\label{proof:polar}
The stability criteria \ref{eq:cond1} and \ref{eq:cond2} both contain terms with edge two. We would like to use the symmetries of the problem to fix this shared edge and reduce the number of free variables. The problem has rotation and reflection symmetry as well as arbitrary edge labels. 

Let us look at an arbitrary pair of non-adjacent edges $\bm{l}_i$ and $\bm{l}_j$ (fig.~\ref{fig:symmetry}, top). Either $\angle(l_i l_j)_{l_k}\leq\pi$ or $\angle (l_i l_j) _{l_m} \leq \pi$, since the two angles together make up a full circle.  Without loss of generality, label the edges so that $\angle(l_i l_j)_{l_k}\leq\pi$. We can then use the problem's rotational symmetry to impose $\hb{l}_k=\hb{x}$, and reflection symmetry to impose that $\rho_k(\hb{z}\times\hb{l}_k)$ lies on the positive y-axis, as shown in fig. \ref{fig:symmetry}. This implies that $p_k$ is positive in the new coordinate system, a fact that will be useful later. We are free to relabel $\bm{l}_k$ as $\bm{l}_2$ and to relabel the rest of the edges in order clockwise from 1 to 4. Since we can perform this procedure starting from any pair of non-adjacent edges $\bm{l}_i$ and $\bm{l}_j$, our arguments in the remainder of this section hold for all pairs of non-adjacent edges.  

\subsection{Bounds on the angle between non adjacent edges}
\label{proof:calc}
\begin{figure}[]
		\centering
    \includegraphics[width=0.5\textwidth]{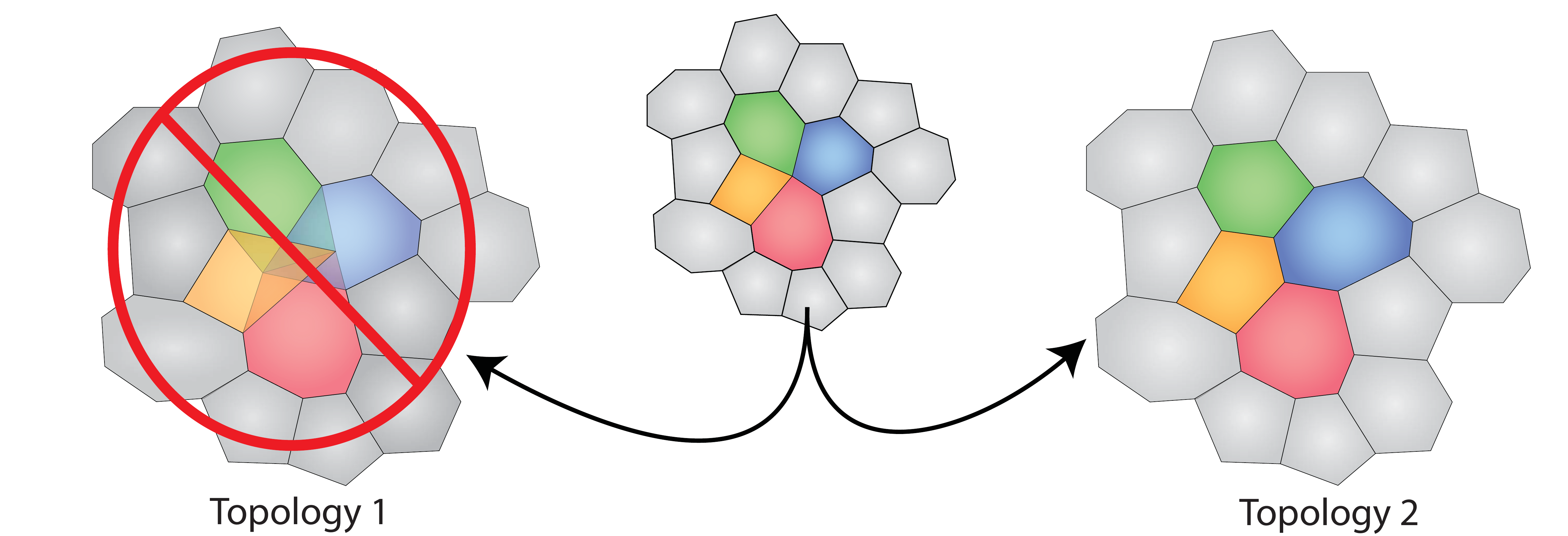}
    \caption{
		Cartoon of the unphysical resolution of a fourfold vertex due to large pressure effects. In the left topology the two resulting threefold vertices are pushed through each other by the pressure of the neighboring cells. This creates a physically impossible state in which cells overlap.
		}
    \label{fig:newedge}
\end{figure}

We next turn to the central problem of determining the implications of the stability criteria of eqs. \ref{eq:cond1} and \ref{eq:cond2} for the angles between edges.  Let $\theta_i$ be the signed angle between an edge and the x-axis, where $\theta_1$ is positive and $\theta_3$ is negative due to the clockwise labeling of edges as shown in fig. \ref{fig:symmetry}. The stability conditions can be written in terms of the $\theta_i$ and $\rho_i$ as
\begin{align}
	1 &\geq 2+\rho_{1}^{2}+\rho_{2}^{2}+2(1+\rho_1 \rho_2)\cos{\theta_1}+2(\rho_2-\rho_1)\sin{\theta_1} \label{eq:compt1} \\
	1 &\geq 2+\rho_3^{2}+\rho_2^{2}+2(1+\rho_3 \rho_2)\cos{\theta_3}+2(\rho_2-\rho_3)\sin{\theta_3}. \label{eq:compt2}
\end{align}
Our goal is to put a lower bound on the angle $\angle (l_1 l_3)_{l_2}=\theta_1-\theta_3$. An important property of our system of inequalities is that the conditions on $\theta_1$ are completely independent of the value of $\theta_3$ and vice versa. Neither variable depends on the other, but they both depend on $\rho_2$. This allows us to break the overall optimization problem of finding the minimum value of $\theta_1-\theta_3$ into two separate sub-problems: finding the minimum value of $\theta_1$ as a function of $\rho_2$ and finding the maximum value of $\theta_3$ as a function of $\rho_2$.

For our first optimization problem, we would like to find the minimum value of $\theta_1$ that can be obtained by varying $\rho_1$ for an arbitrary, fixed value of $\rho_2$ and subject to the constraint of eq. \ref{eq:compt1}. Due to the inequality constraint we cannot use the method of Lagrange multipliers to solve this optimization problem.  Instead, we use its generalization to the case where the optimum can occur either on the boundary of a region or within that region, the Karush-Kuhn-Tucker conditions \cite{Boyd:KKT}. Let the function to be maximized be $h(\theta_1,\rho_1)=-\theta_1$ and the constraining function be $g(\theta_1,\rho_1)=1+\rho_{1}^{2}+\rho_{2}^{2}+2(1+\rho_1 \rho_2)\cos{\theta_1}+2(\rho_2-\rho_1)\sin{\theta_1} \leq 0$. The optimality conditions are then
\begin{align}
	\nabla h(\theta_1,\rho_1) - \lambda \nabla g(\theta_1,\rho_1) &= 0 \label{eq:KKTbeg}\\
	\lambda[g(\theta_1,\rho_1)-0] &=0 \\
	g(\theta_1,\rho_1) &\leq 0 \\
	\lambda &\geq 0 , \label{eq:KKTend}
\end{align}
which produces the system of equations
\begin{align}
	0&=-1+2\lambda\left[(1+\rho_1\rho_2)\sin{\theta_1}+(\rho_1-\rho_2)\cos{\theta_1})  \right] \\
	0&=-2\lambda \left( \rho_1+\rho_2\cos{\theta_1}-\sin{\theta_1} \right) \\
	0&=\lambda g(\theta_1,\rho_1) \\
	0 &\geq g(\theta_1,\rho_1) \\
	0& \leq \lambda 
\end{align}
In section \ref{proof:polar}, we showed that we can used symmetry operations to make $\rho_2$ positive without loss of generality. We also chose to focus on the smaller of the two angles between a pair of non-adjacent edges, so that $\angle (l_1l_3)_{l_2} \leq \pi$, and we numbered the edges clockwise as show in fig. \ref{fig:symmetry} (bottom). This gives additional constraints on the solution: 
\begin{align}
	0 &\leq \rho_2, \\
	0 &\leq \theta_1 \leq \pi.
\end{align}
The solution to the full system of equations is
\begin{align}
	\theta_1 &= \arctan \left[-\rho_2, 1 \right] \label{eq:sol4}.
\end{align}
where $\arctan\left[x,y\right]$ is the angle whose tangent is $y/x$ and that lies the quadrant is given by the signs of $x$ and $y$. 

We may now independently optimize $\theta_3$ for an arbitrary value of $\rho_2$. Let the function to be maximized be $h(\theta_3,\rho_3)=\theta_3$ and the constraining function be $g(\theta_3,\rho_3)=1+\rho_{3}^{2}+\rho_{2}^{2}+2(1+\rho_3 \rho_2)\cos{\theta_3}+2(\rho_2-\rho_3)\sin{\theta_3} \leq 0$. The optimality conditions are the same as eqs. \ref{eq:KKTbeg}-\ref{eq:KKTend}, which produces the system of equations:
\begin{align}
	0&=1+2\lambda \left[(1+\rho_3\rho_2)\sin{\theta_3}+(\rho_3-\rho_2)\cos{\theta_3} \right] \\
	0&=-2\lambda \left( \rho_3+\rho_2\cos{\theta_3}-\sin{\theta_3} \right) \\
	0&=\lambda g(\theta_3,\rho_3) \\
	0 &\geq g(\theta_3,\rho_3) \\
	0& \leq \lambda 
\end{align}
We have an additional two constraints given by the way we set up the problem: 
\begin{align}
	0 &\leq \rho_2, \\
	0 &\geq \theta_3 \geq -\pi.
\end{align}
This system of equations has two solutions:
\begin{align}
	\theta_3 &= \arctan \left[\frac{-2-\rho_2\sqrt{\rho_2^2-3}}{1+\rho_2^2},\frac{-2\rho_2+\sqrt{\rho_2^2-3}}{1+\rho_2^2}\right], \label{eq:solu1} \\
	\theta_3 &= \arctan \left[\rho_2, -1 \right]. \label{eq:solu3}
\end{align}
For all $\rho_2 \geq 0$, the solution of eq. \ref{eq:solu3} is greater than that of eq. \ref{eq:solu1}, so the true maximum is eq. \ref{eq:solu3}. 
Subtracting our two independently optimized solutions we have that the minimum possible value of the angle $\angle(l_1 l_3)_{l_2}$ is
\begin{align}
	\angle(l_1 l_3)_{l_2} &=\theta_1-\theta_3 \notag \\
	&\geq \arctan \left[-\rho_2, 1 \right] - \arctan \left[\rho_2,-1\right] \notag \\
	&= \pi,
\end{align}
where the last identity holds for all positive $\rho_2$.  As we began by choosing $\angle(l_1 l_3)_{l_2}\leq \pi$, either $\angle(l_1 l_3)_{l_2}=\pi$ or the fourfold vertex is unstable.  Moreover, because $\angle(l_1 l_3)_{l_2} + \angle(l_1 l_3)_{l_4} = 2 \pi$, stability then also implies that $\angle(l_1 l_3)_{l_4} = \pi$.  The same holds for any pair of nonadjacent edges, by the argument in section \ref{proof:polar}.  In other words, the fourfold vertex is unstable unless $\hb{l}_1=-\hb{l}_3$ and $\hb{l}_2=-\hb{l}_4$. In the next section, we show that under these assumptions it is impossible to satisfy all three stability conditions \ref{eq:cond1}-\ref{eq:cond3}. 

\subsection{Finding a contradiction when non-adjacent edges have $180^\circ$ separation}
\label{sec:contradiction}
Suppose that $\hb{l}_1=-\hb{l}_3$ and $\hb{l}_2=-\hb{l}_4$.  It is easy to show that condition \ref{eq:cond3} (mechanical equilibrium) is then only satisfied when $\rho_1=\rho_3$ and $\rho_2=\rho_4$. Since $\angle(l_1 l_3)_{l_2} = \theta_1 - \theta_3 = \pi$, $\theta_3 = \theta_1 - \pi$. Substituting this equality and $\rho_3=\rho_1$ into eqs. \ref{eq:compt1} and \ref{eq:compt2} yields
\begin{align}
0 &\geq 1+\rho_{1}^{2}+\rho_{2}^{2}+2(1+\rho_1 \rho_2)\cos{\theta_1}+2(\rho_2-\rho_1)\sin{\theta_1}
\\ 0 &\geq 1 +\rho_1^{2}+\rho_2^{2}-2(1+\rho_1 \rho_2)\cos{\theta_1}-2(\rho_2-\rho_1)\sin{\theta_1}.
\end{align}
Together these two conditions imply 
\begin{equation}
0 \geq 1+\rho_1^2+\rho_2^2,
\end{equation}
which is a contradiction because the right-hand side is always greater than one. Thus, there can be no stable, stationary fourfold vertices in the equal tension vertex model. 

Before moving on from the stationary, equal tension case, we should finally note that, strictly speaking, our proof of instability applies to a vertex model that allows cell overlap.  Although such a situation is not common in practice, it can occur that pressure differences between cells are large enough that they overwhelm the tensions and cause the fourfold vertex to try to resolve by pushing the cells through each other as shown in fig. \ref{fig:newedge}.  If such resolution with overlap is forbidden, the vertex's stability increases, and we cannot at the moment rigorously rule out the possibility that in this case fourfold vertices could become stable in the equal tension model.  In reality, of course, if cell overlap is a concern then there is a good chance the model is being studied in a pathological parameter regime.

\section{Examples of modifications that allow for stable fourfold vertices}
\label{sec:stabelex}
\begin{figure*}[]
		\centering
    \includegraphics[width=0.9\textwidth]{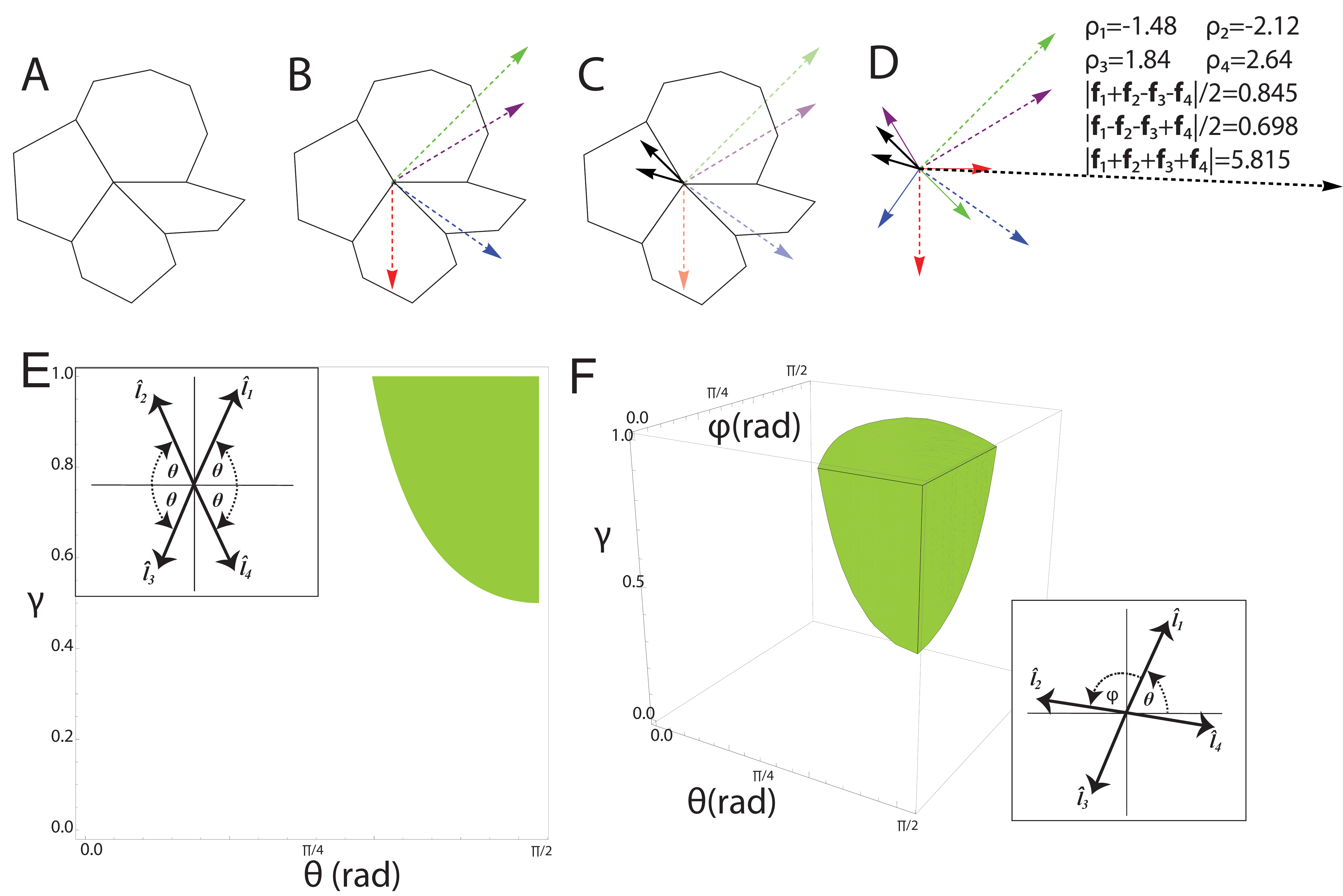}
    \caption{Situations in which fourfold vertices can become stable. \newline
    $\bf{A}$--$\bf{B:}$ Example of a fourfold vertex stabilized through movement. All $\Gamma_i = \Gamma =1$, so that the tension force from each edge is 1.  The vertex is then stable for the quoted values of the pressure differences  \newline
    $\bf{A:}$ The fourfold vertex and adjacent cells. \newline
    $\bf{B:}$ The force from the pressure differences across each edge $\rho_i(\hb{z}\times\hb{l}_i)$ is shown as a dashed line. The magnitudes are to scale.\newline
    $\bf{C:}$ Solid black arrows represent the two values of $\mathcal{F}$ corresponding to the two possible resolution topologies.\newline
    $\bf{D:}$ All of the forces on the vertex are shown. Solid colored arrows represent the edges, which contribute a force of $\hb{l}_i$. The dashed colored arrows represent the force from the pressure across each edge $\rho_i(\hb{z}\times\hb{l}_i)$. Solid black arrows are the two values of $\mathcal{F}$, and the dashed black arrow is the velocity vector. Values of $\rho_i$, $\mathcal{F}$ and the total force are given on the right.  Note that both solid black arrows are shorter than the four arrows giving the edge tensions, indicating that $|\mathcal{F}| < \Gamma < 2 \Gamma$, amply satisfying the stability conditions of eqs. \ref{eq:stability}--\ref{eq:c2}. \newline
		$\bf{E-F:}$ Parameter space in which fourfold vertices with anisotropic edge tensions are stable. \newline
    $\bf{E:}$ Stability for symmetric vertices; $\gamma$ gives the strength of the anisotropy in the tension and $\theta$ gives the angle of the edges with respect to the x-axis (\textit{inset}).  The region of parameter space in which fourfold vertices are stable is shown in green. \newline
    $\bf{F:}$ Stability for asymmetric vertices with paired edges; $\gamma$ gives the strength of the anisotropy in the tension, $\theta$ gives the angle of $\hb{l}_1$ with respect to the $x$ axis, and $\varphi$ gives the angle between $\hb{l}_1$ and $\hb{l}_2$ (\textit{inset}). The region of parameter space in which fourfold vertices are stable is shown in green. 
		}
    \label{fig:example}
\end{figure*}
It was already known that Plateau's model of soap foams, on which the vertex model is based, does not allow for stable fourfold vertices. In the last section we gave a proof that, even with the addition of pressure effects which arise in vertex models with straight edges, there are still no stable fourfold vertices. Given that fourfold vertices are seen in various epithelial tissues \cite{Honda:Checkerb,DiNardo:4fold,Bellaïche:4fold,Maini:4fold,Zallen:Square}, one might naturally wonder what extensions of the model would allow stable fourfold vertices to form.  One well-studied example occurs in the avian oviduct epithelium, where two different types of cells are arranged in a checkerboard pattern with edges between like cell types having higher tension \cite{Honda:Checkerb}.  In this section we will give two examples of modifications which allow for stable fourfold vertices in epithelia even when only a single cell type is present. This gives us some insight into what additional biological mechanisms might exist in epithelia which are not present in simple foams and which could lead to higher order vertices. 
\subsection{Vertices not in mechanical equilibrium}
\label{sec:nonequilibr}
So far we have only considered fourfold vertices which are in mechanical equilibrium. If the vertex is moving relative to the epithelial tissue, eq. \ref{eq:c3general} no longer holds, and the forces associated with the four edges can become very unbalanced.  It turns out that the vertex model then does admit stable fourfold vertices. An example of such a stable state is given in fig. \ref{fig:example}. The observation that moving fourfold vertices tend to be more stable than their stationary counterparts might explain why they have been observed to persist in tissues undergoing rapid morphogenetic movements~\cite{Campinho:4fold,Maini:4fold}.

\subsection{Anisotropic tension}
In previous sections, we investigated vertex stability in a model in which all edges have the same tension.  Unlike foams, however, cells can regulate their tensions so that these differ from one edge to the next. One example of this is the anisotropic edge tensions produced through the planar cell polarity pathway \cite{Axelrod:PCP,Munro:PCP,Bellaiche:PCP,Eaton:PCP,Mao:PCP,Assémat:PCP,Zallen:Square} (which breaks rotational symmetry by defining a preferred direction in the plane of the epithelium). 

A very simple model of planar cell polarity is to assume that tension regulating proteins (such as myosin) are recruited to edges based on the edges' angle with the overall polarity orientation, so that edges have an anisotropic tension given (in appropriate dimensionless units) by $\Gamma_i=1+\gamma \cos{2 \theta_i}$, where $\gamma \in [0,1]$ gives the strength of the anisotropy, and $\theta_i$ is the angle between the edge and the planar polarity axis (which we will always take to be the $x$ axis) \cite{Mao:PCP,Glazier:PCP,Rauzi:PCP}. We will make the further assumption that there is some time lag for proteins to move onto the newly forming edge, so that the new edge tension will not depend on the angle, but instead will simply be the average tension $\Gamma_\delta=1$. In order to further simplify the model we will also assume that effects from pressure are negligible. The force on a fourfold vertex is then described by five variables: $\gamma$ and the four angles $\theta_i$ between the edges and the polarity axis.

With the additional effects of polarization some stable fourfold states exist. We begin our examination of the stable states by looking only at states which are symmetric about both the $x$ and $y$ axes (fig. \ref{fig:example}E, inset). Let $\theta$ be the angle between the high tension x-axis and the edges. From the conditions given in eqs. \ref{eq:c1general}-\ref{eq:c2general} it is easy to show that the vertex is stable if it satisfies both:
\begin{align}
  1 &>2(1+\gamma \cos 2\theta )\cos \theta  \\ 
  1 &>2(1+\gamma \cos 2\theta )\sin \theta.  
\end{align}
The solutions to this series of inequalities are shown in fig. \ref{fig:example}E. In general we have stable fourfold vertices when the strength of the polarization is fairly high and $\theta$ is near $\frac{\pi}{2}$. This makes intuitive sense because this represents all of the edges being near the low tension
axis and the strength of the tension being relatively low. 

We now lift the restriction of symmetry in order to look for more general instances of stability. We will assume that the edges come in equal and opposite pairs ($\hb{l}_1=-\hb{l}_3$  and $\hb{l}_2=-\hb{l}_4$), so that mechanical equilibrium is ensured and the number of free parameters is still low. (With this restriction, we still cannot explore all possible states of the model, but the variety of available vertex geometries is large enough to clearly demonstrate how polarized tensions can lead to stability.) We now have three free parameters. Let $\theta$ be the angle between the high tension axis (the $x$-axis) and the first edge, $\varphi$ be the angle between the first and second edges, and $\gamma$ be the strength of the polarization. We then have that $\theta_1=\theta$, $\theta_2=\theta+\varphi$, $\theta_3=\theta+\pi$, and $\theta_4=\theta+\varphi+\pi$. In order to have stability the following two inequalities must hold. 
\begin{align}
   1 &> \bigg((1+\gamma \cos 2\theta )\cos \theta 													\notag \\
   &+\big[1+\gamma \cos 2(\theta +\phi )\big]\cos(\theta +\phi ) \bigg)^2  						\notag \\
  &+ \bigg( (1+\gamma \cos 2\theta )\sin \theta 															\notag \\
  &+\big[1+\gamma \cos 2(\theta +\phi )\big]\sin (\theta +\phi ) \bigg)^2 										\\
  1 & > \bigg( (1+\gamma \cos 2\theta )\cos \theta 													\notag \\
  &+\big[1+\gamma \cos 2(\theta +\phi -\pi )\big]\cos (\theta +\phi -\pi )\bigg)^2 	\notag \\
  &+ \bigg((1+\gamma \cos 2\theta )\sin \theta 															\notag \\
  &+\big[1+\gamma \cos 2(\theta +\phi -\pi )\big]\sin (\theta +\phi -\pi )\bigg)^2
\end{align}
The solution to this series of inequalities is shown in fig. \ref{fig:example}F. This is reasonable because more angles are stable as the amount of polarization increases and once again these angles represent the edges placed near the low tension axis. 

Stable fourfold vertices are seen in some systems with planar cell polarity \cite{DiNardo:4fold,Bellaïche:4fold,Zallen:Square}. The stability of these vertices may be due to the decreased tension on edges along the low tension axis.

\section{Implications for computational models}
\label{sec:computation}

Although vertex models are widely used to simulate epithelial dynamics, there is currently no standard procedure for dealing with T1 transitions in such simulations. Some naive implementations can resolve fourfold vertices in ways that produce unphysical behavior.  For example, approaches that automatically perform a T1 transition whenever an edge becomes too small, or more generally that assume that a fourfold vertex must always break up into two threefold vertices, can lead to spurious oscillations when the fourfold vertex should in fact be stable; importantly, as we showed in the preceding section, moving vertices can become stable even when all tensions are equal, so this issue can in principle arise in almost all vertex model simulations.  Something similar can occur when a fourfold vertex is resolved into two threefold vertices with a separation $\bm{l}_{\delta}$ that is not parallel to $\bm{\mathcal{F}}$ (though this phenomenon can be avoided---see below---if $|\bm{l}_{\delta}|$ is chosen small enough).
 In this section we briefly describe a method, based on the theoretical developments of the previous sections, that carries out T1 transitions in a consistent fashion and so avoids these and similar difficulties.  Complete pseudo-code for this algorithm appears in the appendix. 

The essential idea of our algorithm is that T1 transitions must be dealt with in two steps:  First, an edge whose length is below a chosen cutoff is removed
and the two threefold vertices joined by that edge are merged into a single fourfold vertex.  Then, one checks the stability of the fourfold vertex against breaking in both allowed topologies (recognizing, as shown in Sec. \ref{sec:stabelex}, that the fourfold vertex could be stable).  This requires creating temporary threefold vertices, with zero separation, and corresponding edges, so that the forces on the two new vertices can be calculated in each topology.  Depending on the stability of the fourfold vertex, three outcomes are possible:  1)  The fourfold vertex is found to be stable and allowed to persist.  (In this case, the vertex could still become unstable at some later time, so one must continue to monitor its stability as the simulation progresses.)  2)  The fourfold vertex resolves into two threefold vertices in the same topology as the original threefold vertices.  One thus effectively rejects the proposed T1 transition even though the initial edge length is less than the cutoff.  3)  The fourfold vertex resolves into two threefold vertices in the new topology, and a T1 transition occurs.

Once it has been determined that a fourfold vertex is unstable one needs to make a new edge of finite length, which raises the question of the most appropriate orientation for the new edge. From eq. \ref{eq:theta} the new edge rotates at a rate
\begin{equation}
\dot{\theta}=- \frac{\mathcal{F}}{\mu l_\delta}\sin\theta,
\end{equation}
and its length changes according to
\begin{equation}
\dot{l}_\delta =  \frac{\mathcal{F}}{\mu} \cos{\theta} - \frac{2 \Gamma_{\delta}}{\mu}.
\end{equation}
The edge orientation must clearly relax to $\theta=0$ as long as $\bm{\mathcal{F}}$ remains approximately constant over the relaxation timescale. Because $\dot{\theta}$ diverges like $1/l_\delta$, it is reasonable to guess that this will be the case if the initial edge length $l_{0\delta}$ is chosen small enough.  More precisely, one can estimate that the edge relaxes to $\theta=0$ on a timescale $\frac{\mathcal{F}}{\mu l_{0\delta}}$.  Over that time, the change in edge length will be of order $l_{0\delta}$.  Thus, the fractional change in the new edge's length during the relaxation process is of order one.  Nonetheless, if $l_{0\delta}$ is small compared to the scale, typically of order a cell size, over which $\bm{\mathcal{F}}$ changes appreciably, then the variation in $\bm{\mathcal{F}}$ over the time it takes $\theta$ to rotate to zero can still be neglected.  We thus conclude that if they are short enough, new edges will always quickly rotate to become parallel with $\bm{\mathcal{F}}$, whatever their initial orientation.  It is then reasonable in simulations simply always to create new edges with $\theta = 0$.

\section{Discussion}
Vertex models are important tools to study the interplay between local cell mechanics and global tissue shape and motion.  One aspect of this interaction during tissue remodeling and development is the T1 transition, in which a fourfold vertex is formed as an intermediary stage.  More generally, the local behavior of fourfold vertices affects cell shape and mechanics, and thereby morphogenesis at larger scales. 

Here, we have introduced a formulation of the stability of fourfold vertices in vertex models with straight edges that holds for arbitrary edge tensions and cell pressures (whether or not derived from an underlying energy function).  Using this formulation, we have given the first proof that, in the simplest case of equal edge tensions and vertices in mechanical equilibrium---analogous to the conditions in a dry foam---fourfold vertices are never stable in vertex models, just as they are not in the Plateau model of foams.  

We have also shown that if either of the assumptions of equal edge tensions or mechanical equilibrium is relaxed, fourfold vertices can become stable.  Interestingly, long-lived fourfold and higher order vertices have been observed in epithelia moving relative to the surrounding fluid \cite{Maini:4fold,Campinho:4fold} and in tissues where junctional tensions are influenced by planar cell polarity \cite{DiNardo:4fold,Bellaïche:4fold,Zallen:Square}, suggesting that both stabilization scenarios may have biological relevance. 

Lastly, our treatment of vertex stability has clear implications for the simulation of vertex models and especially for the implementation of T1 transitions in computational modeling (see appendix \ref{appendix}).  Moreover, whereas our analytic results apply to models that in principle allow for cell overlap, in computational formulations this problem can be addressed by checking for overlap after T1 transitions.  Disallowing overlap may stabilize some fourfold vertices in the limit where the force on the vertex from the cell pressure dominates over the tension on the edges (though such parameter regimes are not those thought to be physically relevant in most studies of vertex models, and in particular one could question whether it is a good approximation to force edges to remain straight when pressures are high enough). Our discussion in this paper has been limited to fourfold vertices, but higher order vertices, like the rosettes seen during \textit{Drosophila} germband extension \cite{Zallen:T1example}, can be investigated in an entirely analogous manner, by checking whether the vertex is stable against breaking up into every possible combination of two lower order vertices; of course, the number of stability conditions will increase rapidly with the order of the vertex.

Although the relatively simple models for determining pressures and edge tensions that we have adopted here capture many aspects of the behavior of real epithelia, certain systems clearly require more sophisticated descriptions.  For example, in the pupal dorsal notum of \textit{Drosophila} \textit{pten} mutants, vertices are seen to undergo oscillatory T1 transitions that appear to be driven by disparities in the timescales for transport of different proteins to newly formed edges \cite{Bellaïche:4fold,Mao:oss}.  Our description of vertex stability can readily be extended to include many effects along these lines.  In particular, as long as the new edge is much shorter than the existing edges, the stability problem can still be expressed in terms of the dynamics of the new edge $\bm{l}_\delta$, which in turn are determined by the---now possibly time-dependent---tensions and pressures of the surrounding edges and cells.  Similarly, our formalism can encompass buckling of the epithelial sheet into the third dimension~\cite{Reuter:VM,Du:VM} without any significant modifications, because even a bent epithelium appears locally flat when $l_\delta$ is much less than the sheet's radius of curvature, as it must be immediately after a fourfold vertex has broken up. 

On the other hand, our formalism assumes that vertex stability is solely a consequence of local edge tensions and cell pressures; it does not include the effects of other phenomena that might be relevant in some biological systems and that would require more substantial changes to our basic model. For example, it is possible that in some circumstances cells could recruit proteins specifically to fourfold vertices to stabilize or destabilize them.  Similarly, the models studied here neglect effects associated with the fluid dynamics of molecular transport to and from vertices \cite{Stone:T1} and include interactions between the epithelium and its substrate only in the coarsest fashion, as one of the sources of the local friction force on vertices.   Our calculations thus represent only an initial step towards understanding the rich physics of topology changes and vertex stability in epithelia and planar foams.

\section{Acknowledgments}
This material is based upon work supported by the National Science Foundation under Grant No. DMR-1056456 and an NSF Graduate Research Fellowship under Grant No. DGE-1256260. 
\vspace{3 mm}

\noindent
Author Contributions: 

Conceptualization: MAS ZJ DKL

Formal Analysis: MAS ZJ

Writing: MAS DKL

\vspace{3 mm}

\noindent
* \textit{Present address}: Department of Mechanical Engineering and Applied Mechanics, University of Pennsylvania, Philadelphia, PA 19104.

\FloatBarrier
\bibliographystyle{ieeetr}
\bibliography{master3}

\appendix
\onecolumn
\SetKwComment{Comment}{$\triangleright$\ }{}
\newcommand{\OR}{\KwSty{ or }}
\newcommand{\AND}{\KwSty{ and }}

\section{Pseudo-code for T1 transitions}
\label{appendix}
Algorithms 1-4 give pseudo-code implementing T1 transitions as described in section~\ref{sec:computation}. The code assumes an object oriented language (such as C++ or Java) with cell, edge, and vertex objects already defined. We will assume that the edges store data on their neighboring vertices and cells. The cells and vertices only store data on their neighboring edges, and functions have been written to get the other neighboring objects if needed. Objects are referred to in C++ style so that someobject.somedata refers to the data \textit{somedata} stored by the object \textit{someobject}.

The function T1 takes a small edge and replaces it with a new fourfold vertex, and then calls the function ResolveFourfoldVertex on the new vertex.

The function ResolveFourfoldVertex takes as input a fourfold vertex. It calls CheckStability on each of the possible resolution topologies to determine their stability. Once the correct resolution topology has been established the function calls BreakFourfoldVertex to update the edges, cells, and vertices involved in the T1 transition.

The function CheckStability takes as input a fourfold vertex and its associated edges and cells. The cells and edges must be given in clockwise order. The function will create temporary objects representing breaking the fourfold vertex such that edges $e_1$ and $e_2$ share a common vertex. The force $\mathcal{F}$ is calculated and returned, and the temporary objects are deleted.

The function BreakFourfoldVertex takes a resolution topology for a fourfold vertex as input and creates the new edge and correctly resigns the neighboring edges, vertices, and cells in the new topology. 

(Including both the CheckStability and BreakFourfoldVertex functions may seem redundant, but it is vital to have both to deal with the rare but possible case in which a fourfold vertex is unstable to breaking up in both topologies. In this case the vertex should break in the topology in which it is most unstable.)

\begin{algorithm}
\caption{T1}
	\KwIn{$e_0$ the edge to undergo T1}
	\KwOut{None. The function will update the effected edges cell and vertices so they are in the correct locations and have the correct neighbors following a T1 transition.}
$v1 \gets e0.vertex1$\;
$v2 \gets e0.vertex2$\;
$c1 \gets e0.cell1$\;
$c2 \gets e0.cell2$\;
\Comment{Do not T1 edges which neighbor triangles. This would produce cells with only two sides.}
\If{$c1.EdgeNumber \leq 3 \OR c2.EdgeNumber\leq 3$}{EXIT}
\Comment{Get the edges which will make the fourfold vertex}
\ForAll{$e \in \{v1.edges \OR v2.edges \}$}{
	\If{$e \neq e0$}{
		$list4foldedges \gets e$ 			
		}
}
\Comment{set the next position of the two vertices to the center of the edge}
$v1.xnext \gets e0.center$  \;	
$v2.xnext \gets e0.center$ \;	
\Comment{move vertices updating any periodic boundary flags if nessacary} 
\FuncSty{MoveVertex}{(v1)} \;							 
\FuncSty{MoveVertex}{(v2)} \;		
\Comment{make the new vertex} 			
$vnew \gets e0.center$\;					
\ForAll{$e \in list4edges$}{
		$vnew \gets e$ 		
	}
\Comment{delete two old vertices} 
\FuncSty{Delete}{(v1)}\;						
\FuncSty{Delete}{(v2)}\; 
\Comment{Remove e0 from the list of edges in its two neighboring cells}
\For{$c \in \{c1,c2\}$}{
	\ForAll{ $e \in c$}{
		\If{$e=e0$}{
			remove $e$ \;
		}
	}
}
\FuncSty{Delete}{(e0)}								\Comment*[r]{delete the central edge e0}
\FuncSty{ResolveFourfoldVertex}{(vnew)}							\Comment*[r]{Resolve the fourfold vertex}
\end{algorithm}

\begin{algorithm}
\caption{{\sc ResolveFourfoldVertex}}
\DontPrintSemicolon 
\KwIn{$v0$ the fourfold vertex to resolve}
\KwOut{None}
\Comment{Make lists of the edges and cells in clockwise order}
$e \gets v.CWEdges$\;
$c \gets v.CWCells$\; 
\Comment{find the stability of each configuration}
$f1 \gets $ \FuncSty{CheckStability}(v, e[0], e[1], e[2], e[3], c[0], c[1], c[2], c[3])\; 
$f2 \gets $ \FuncSty{CheckStability}(v, e[3], e[0], e[1], e[2], c[3], c[0], c[1], c[2])\; 
\Case{f1=0 \AND f2=0}{
	EXIT \Comment*[r]{The vertex is stable so exit}
}
\Case{$f1 \geq f2$}{
	\Comment{The vertex is unstable and should resolve in the first topology}
	\FuncSty{BreakFourfoldVertex}(v, e[0], e[1], e[2], e[3], c[0], c[1], c[2], c[3])
}
\Case{$f2>f1$}{
	\Comment{The vertex is unstable and should resolve in the second topology}
	\FuncSty{BreakFourfoldVertex}(v, e[3], e[0], e[1], e[2], c[3], c[0], c[1], c[2])
}
\end{algorithm}

\begin{algorithm}
\caption{{\sc CheckStability}}
\DontPrintSemicolon 
\KwIn{ v, e1, e2, e3, e4, c1, c2, c3, c4\;
v: the fourfold vertex\;
			e1, e2, e3, e4: the four edges of v in clockwise order such that (e1,e2) will be neighbors and (e3,e4) will be neighbors when the vertex is split\;
			c1, c2, c3, c4: The four cells of v in clockwise order such that c1 has edges e1, and e2. 
}
\KwOut{Creates temporary objects representing the vertex splitting such that edges (e1,e2) and (e3,e4) are paired and cells c2 and c4 are neighbors. It returns the magnitude of the force pulling the vertices apart. If the vertex is stable against breaking in this topology it returns 0.}
\FuncSty{CoppyAll} $v'\gets v$, $e1' \gets e1$, $c1' \gets c1, ...$ \Comment*[r]{Make temporary objects} 

\Comment{Make the new edge (enew) and vertices (v12, and v34) resulting from the split into two threefold vertices}						
	$v12 \gets v.x$	\Comment*[r]{the vertex on edges e1' and e2'}	
	$v12 \gets \{enew, e1', e2' \}$			\;		
	$v34 \gets v.x$		\Comment*[r]{the vertex on edges e3' and e4'}	
	$v34 \gets \{enew, e3', e4' \}$				\;			
	$enew.length \gets 0$		\;	
	$enew \gets \{v12, v34 \}$ \;
	$enew \gets \{c2', c4' \}$ \;
\For{$e \in \{e1', e2', e3', e4' \}$}{
		$e$ delete $v$ 		\Comment*[r]{Update the four edges}
		\If{ $e \in \{e1',e2'\}$}{
			$e \gets v12$
		}
		\Else{
			$e \gets v34$
		}
}
$c2' \gets enew$  \Comment*[r]{Update the cells}
$c4' \gets enew$ \;
\Comment{Calculate $\mathcal{F}$ as given in Sect3A. Let e.FindForce(v) return the force on vertex v from edge e given by $\Gamma_e \hat{\bf{l}_e} + \frac{p_e}{2}(\hat{\bf{z}}\times \bf{l}_e)$. }
$\mathcal{F} \gets ( e1.\FuncSty{FindForce}(v12) + e2.\FuncSty{FindForce}(v12) + e3.\FuncSty{FindForce}(v34) + e4.\FuncSty{FindForce}(v34) )/2$ \;
\If{$\mathcal{F} > enew.tension$}{
	\Return magnitude($\mathcal{F}$) \; 
}
\Else{
	\Return 0; 
}	
\end{algorithm}

\begin{algorithm}
\caption{{\sc BreakFourfoldVertex}}
\DontPrintSemicolon 
\KwIn{ v, e1, e2, e3, e4, c1, c2, c3, c4\;
v: the fourfold vertex\;
			e1, e2, e3, e4: the four edges of v in clockwise order such that (e1,e2) will be neighbors and (e3,e4) will be neighbors when the vertex is split\;
			c1, c2, c3, c4: The four cells of v in clockwise order such that c1 has edges e1, and e2. 
}
\KwOut{None}
\Comment{Make the new edge (enew) and vertices (v12, and v34) resulting from the split into two threefold vertices}						
	$v12 \gets v.x+(\frac{L}{2}\hat{\mathcal{F}}$)	\Comment*[r]{Where L specifies new edge lengths}	
	$v12 \gets \{enew, e1', e2' \}$				\;		
	$v34 \gets v.x-(\frac{L}{2}\hat{\mathcal{F}}$)	\Comment*[r]{Where L specifies new edge lengths}	
	$v34 \gets \{enew, e3', e4' \}$					\;	
	$enew.length \gets L$		\;	
	$enew \gets \{v12, v34 \}$ \;
	$enew \gets \{c2, c4 \}$ \;
\For{$e \in \{e1, e2, e3, e4 \}$}{
		$e$ delete $v$ 		\Comment*[r]{Update the four edges}
		\If{ $e \in \{e1,e2\}$}{
			$e \gets v12$
		}
		\Else{
			$e \gets v34$
		}
}
$c2 \gets enew$  \Comment*[r]{Update the cells}
$c4 \gets enew$ \;
\FuncSty{Delete} v \;
\end{algorithm}

\end{document}